\documentclass[a4paper,12pt]{article}
\pdfoutput=1 % if your are submitting a pdflatex (i.e. if you have images in pdf, png or jpg format)
\usepackage{jcappub} % for details on the use of the package, please
\usepackage[symbol]{footmisc}
\usepackage{changes}
\usepackage{amsmath,amssymb}
\usepackage{graphicx}
\usepackage{adjustbox}
\usepackage{multirow}
\usepackage[normalem]{ulem}
\usepackage{bm}

\usepackage{graphicx}   % include figures
\usepackage{tabulary}
\usepackage{color} 

\usepackage{soul} % 导入 soul 包
\usepackage{color, xcolor} % 颜色包, color 必须导入
\setstcolor{red}

\newcommand{\nc}{\newcommand*}

\nc{\al}{\alpha}
\nc{\s}{\sigma}
\nc{\dt}{\delta}
\nc{\Dt}{\Delta}
\nc{\Ld}{\Lambda}
\nc{\p}{\partial}
\nc{\om}{\omega}
\nc{\Om}{\Omega}
\nc{\rd}{\mathrm{d}}
\nc{\Od}[1]{\mathcal{O}(#1)} % order operator
\nc{\kp}{\kappa}

\def\({\left(}
\def\){\right)}
\def\[{\left[}
\def\]{\right]}
\def\e{\begin{equation}}
\def\q{\end{equation}}
\def\m{\begin{eqnarray}}
\def\n{\end{eqnarray}}

\nc{\Eq}[1]{Eq.~\eqref{#1}}     % equation
\nc{\Fig}[1]{Fig.~\ref{#1}}     % figure
\nc{\Table}[1]{Table~\ref{#1}}  % table
\nc{\Sec}[1]{Sec.~\ref{#1}}     % section

\nc{\Msun}{M_\odot}             % solar mass
\nc{\fpbhn}{f_{\mathrm{pbh0}}}    % f_pbh
\nc{\mR}{\mathcal{R}} % merger rate density
\nc{\seq}{\sigma_{\mathrm{eq}}}
\nc{\ogw}{\Omega_{\mathrm{GW}}}
\nc{\gpcyr}{\mathrm{Gpc}^{-3}\,\mathrm{yr}^{-1}}
\nc{\lvc}{LIGO/Virgo} % LIGO-VIRGO collaboration
\nc{\SNR}{\mathrm{SNR}} % signal to noise ratio
\nc{\mmin}{{m_{\mathrm{min}}}}
\nc{\mmax}{{m_{\mathrm{max}}}}
\nc{\Mmin}{{M_{\mathrm{min}}}}
\nc{\fmin}{{f_{\mathrm{min}}}}
\nc{\VT}{\mathrm{VT}}
\nc{\rhoGW}{\rho_{\mathrm{GW}}}
\nc{\vth}{\vec{\theta}}
\nc{\vd}{\vec{d}}
\nc{\vla}{\vec{\lambda}}
\nc{\Nobs}{N_{\mathrm{obs}}}
\nc{\av}[1]{\langle #1 \rangle} % average bracket
\nc{\km}{\mathrm{km}}
\nc{\Mpc}{\mathrm{Mpc}}
\nc{\Tobs}{T_{\mathrm{obs}}}
\nc{\Ntemp}{N_{\mathrm{temp}}}

\nc{\addref}{[\textcolor{red}{add ref}] } % placeholder of references
\nc{\eg}{\textit{e.g.~}}
\nc{\app}{\approx}
\nc{\hf}{\frac{1}{2}}
\nc{\discuss}{\textcolor{red}{Add discussion here!}}

\nc{\mpbh}{m_{\rm{pbh}}}
\nc{\cR}{\mathcal{R}}
\nc{\mU}{{\mathcal{U}}}
\nc{\Mc}{{M_\mathrm{c}}}
\nc{\Mf}{{M_\mathrm{f}}}
\nc{\red}[1]{\textcolor{red}{#1}}
\nc{\yellow}[1]{\textcolor{yellow}{#1}}
\nc{\green}[1]{\textcolor{green}{#1}}
\nc{\blue}[1]{\textcolor{blue}{#1}}
\nc{\fnl}{F_{\mathrm{NL}}}
\nc{\gnl}{G_{\mathrm{NL}}}
\nc{\MG}{\mathcal{M}_{\mathrm{G}}}
\nc{\MNG}{\mathcal{M}_{\mathrm{NG}}}

\begin{document}
	
\title{Exploring the NANOGrav Signal and Planet-mass Primordial Black Holes through Higgs Inflation} 
%%%%%%%%%%%%%%%%%%%%%%%%%%%%%%%%%%%%%%%%%%%%%%%%%%%%%%%%%%%%%%%%%
\author{Zhu~Yi,$^{a,b}$}
\author{Zhi-Qiang~You,$^{c,d,b}$}
\author{You~Wu,$^{e}$}
\author{Zu-Cheng~Chen,\note{Corresponding author.}$^{f,g,d,b,*}$}
\author{Lang~Liu$^{d,b,*}$}

\affiliation{$^a$Faculty of Arts and Sciences, Beijing Normal University, Zhuhai 519087, China}
\affiliation{$^b$Advanced Institute of Natural Sciences, Beijing Normal University, Zhuhai 519087, China}
\affiliation{$^c$Henan Academy of Sciences, Zhengzhou 450046, Henan, China}
\affiliation{$^d$Department of Astronomy, Beijing Normal University, Beijing 100875, China}
\affiliation{$^e$College of Mathematics and Physics, Hunan University of Arts and Science, Changde, 415000, China}
\affiliation{$^f$Department of Physics and Synergetic Innovation Center for Quantum Effects and Applications, Hunan Normal University, Changsha, Hunan 410081, China}
\affiliation{$^g$Institute of Interdisciplinary Studies, Hunan Normal University, Changsha, Hunan 410081, China}

\emailAdd{yz@bnu.edu.cn}	
\emailAdd{you\_zhiqiang@whu.edu.cn}
\emailAdd{youwuphy@gmail.com}	
\emailAdd{zucheng.chen@bnu.edu.cn}
\emailAdd{liulang@bnu.edu.cn}	

\abstract{
The data recently released by the North American Nanohertz Observatory for Gravitational Waves (NANOGrav) provides compelling evidence supporting the existence of a stochastic signal that aligns with a gravitational-wave background. 
We show that the scalar-induced gravitational waves from the Higgs inflation model with the parametric amplification mechanism can explain this signal. Such a gravitational-wave background naturally predicts the substantial existence of planet-mass primordial black holes, which can be planet 9 in our solar system and the lensing objects for the ultrashort-timescale microlensing events observed by the Optical Gravitational Lensing Experiment.  Therefore, the NANOGrav signal, the potential Planet 9 in our solar system, and the Optical Gravitational Lensing Experiment can be explained within the framework of Higgs inflation. 
%The future observations of nanoHertz stochastic gravitational-wave background and planet-mass primordial black holes provide an invaluable opportunity to confirm the Higgs inflation, which connects two fundamental aspects of theoretical physics: particle physics and cosmology.
}
	
\maketitle
%%%%%%%%%%%%%%%%%%%%%%%%%%%%%%%%%%%%%%%%%%%%%%%%%%%%%%%%%%%%%%%%%
\section{Introduction} 	
Subsequent to the detection of gravitational waves (GWs) arising from the coalescence of black holes and neutron stars by LIGO-Virgo-KAGRA~\cite{Abbott:2016blz,TheLIGOScientific:2017qsa,LIGOScientific:2018mvr,LIGOScientific:2020ibl,LIGOScientific:2021usb,LIGOScientific:2021djp}, the forthcoming inspiring discovery may be the identification of the stochastic gravitational-wave background (SGWB), which can span a large frequency range from $10^{-20}$ Hz to $10^{8}$ Hz. The pulsar timing array (PTA) can ascertain the nanoHertz frequency band of the SGWB, which proves to be a valuable window for detecting GWs originating from the early Universe. Recently, the North American Nanohertz Observatory for Gravitational Waves (NANOGrav) \cite{NANOGrav:2023gor, NANOGrav:2023hde},  as well as Parkers PTA (PPTA)~\cite{Zic:2023gta,Reardon:2023gzh},  European PTA (EPTA) along with India PTA (InPTA) \cite{EPTA:2023sfo,Antoniadis:2023ott}, and Chinese PTA (CPTA) \cite{Xu:2023wog}  have reported compelling evidence for a common-spectrum signal consistent with the Hellings-Downs spatial correlations~\cite{Hellings:1983fr}, supporting the existence of a stochastic signal that aligns with an SGWB. While a lot of potential sources exist within the PTA  window~\cite{Li:2019vlb,Vagnozzi:2020gtf,Chen:2021wdo,Wu:2021kmd,Chen:2021ncc,Sakharov:2021dim,Benetti:2021uea,Chen:2022azo,Ashoorioon:2022raz,PPTA:2022eul,Wu:2023pbt,IPTA:2023ero,Wu:2023dnp,Dandoy:2023jot,Madge:2023cak,Chen:2023zkb}, the discernment of whether this signal originates from astrophysical or cosmological phenomena remains the subject of intensive investigation~\cite{NANOGrav:2023hvm,Antoniadis:2023xlr,King:2023cgv,Niu:2023bsr,Bi:2023tib, Liu:2023pau,Vagnozzi:2023lwo,Han:2023olf,Li:2023yaj,Franciolini:2023wjm,Shen:2023pan,Kitajima:2023cek,Franciolini:2023pbf,Addazi:2023jvg,Cai:2023dls,Inomata:2023zup,Murai:2023gkv,Li:2023bxy,Anchordoqui:2023tln,Liu:2023ymk,Abe:2023yrw,Ghosh:2023aum,Figueroa:2023zhu,Yi:2023mbm,Wu:2023hsa,Li:2023tdx,Geller:2023shn,You:2023rmn,Antusch:2023zjk,Ye:2023xyr,HosseiniMansoori:2023mqh,Jin:2023wri,Zhang:2023nrs,ValbusaDallArmi:2023nqn,DeLuca:2023tun,Choudhury:2023kam,Gorji:2023sil,Das:2023nmm,Yi:2023tdk,Ellis:2023oxs,He:2023ado,Balaji:2023ehk,Kawasaki:2023rfx,Cannizzaro:2023mgc,King:2023ayw,Maji:2023fhv,Bhaumik:2023wmw,Zhu:2023lbf,Basilakos:2023xof,Huang:2023chx,Jiang:2023gfe,DiBari:2023upq,Aghaie:2023lan,Garcia-Saenz:2023zue,InternationalPulsarTimingArray:2023mzf}.

One promising explanation for the observed signal is the scalar-induced gravitational waves (SIGWs) generated by the primordial curvature perturbations at small scales~\cite{Ananda:2006af,Baumann:2007zm,Garcia-Bellido:2016dkw,Inomata:2016rbd,Garcia-Bellido:2017aan,Kohri:2018awv,Cai:2018dig,Lu:2019sti,Yuan:2019wwo,Chen:2019xse,Xu:2019bdp,Yuan:2019udt,Cai:2019cdl,Yuan:2019fwv,Yi:2020kmq,Yi:2020cut,Liu:2020oqe,Gao:2020tsa,Yuan:2020iwf,Yuan:2021qgz,Yi:2021lxc,Yi:2022anu,Yi:2022ymw,Yuan:2023ofl,Meng:2022ixx}, which is favored by the NANOGrav data over the supermassive black hole binaries (SMBHBs) scenario through Bayesian analysis~\cite{NANOGrav:2023hvm}. When the primordial curvature perturbations attain considerable magnitudes, they can produce a significant SGWB via the second-order effects arising from the nonlinear coupling of perturbations. Moreover, the emergence of PBHs can be triggered by the presence of large curvature perturbations~\cite{Zeldovich:1967lct,Hawking:1971ei,Carr:1974nx}. In recent years, PBHs have attracted considerable interest~\cite{Belotsky:2014kca,Carr:2016drx,Garcia-Bellido:2017mdw,Carr:2017jsz,Germani:2017bcs,Chen:2018rzo,Liu:2018ess,Chen:2018czv,Liu:2019rnx,Fu:2019ttf,Liu:2019lul,Cai:2019bmk,Chen:2019irf,Liu:2020cds,Fu:2020lob,Liu:2020vsy,Liu:2020bag,Wu:2020drm,DeLuca:2020sae,Vaskonen:2020lbd,DeLuca:2020agl,Domenech:2020ers,Hutsi:2020sol,Chen:2021nxo,Kawai:2021edk,Braglia:2021wwa,Cai:2021wzd,Liu:2021jnw,Braglia:2022icu,Liu:2022wtq,Zheng:2022wqo,Chen:2022qvg,Liu:2022iuf,Chen:2022fda,Inomata:2022yte,Guo:2023hyp,Cai:2023uhc,Meng:2022low} (see also reviews~\cite{Sasaki:2018dmp,Carr:2020gox,Carr:2020xqk}) because they are promising candidates for dark matter~\cite{Sasaki:2018dmp,Carr:2020gox,Carr:2020xqk}. Furthermore, they offer a plausible explanation for the observed binary black holes detected by LIGO-Virgo-KAGRA~\cite{Bird:2016dcv,Sasaki:2016jop}.

To generate significant SIGWs compatible with the recent PTA signal, the amplitude of the primordial curvature power spectrum in the small scales needs to be enhanced by around seven orders of magnitude compared with that in the cosmic microwave background (CMB) scale. One way to amplify the primordial curvature power spectrum is through the parametric amplification mechanism~\cite{Cai:2019bmk} that is independent of the inflation potential. In this work, we show that the recently reported NANOGrav signal can be explained by the SIGWs from the Higgs inflation with the parametric amplification mechanism. Such an SGWB naturally predicts the substantial existence of planet-mass PBHs, which can be planet 9 in our solar system~\cite{Scholtz:2019csj} and the lensing objects for the ultrashort-timescale microlensing events observed by the Optical Gravitational Lensing Experiment (OGLE)~\cite{Mroz:2017mvf, Niikura:2019kqi}.

%%%%%%%%%%%%%%%%%%%%%%%%%%%%%%%%%%%%%%%%%%%%%%%%%%%%%%%%%%%%%%%%%
\section{Parametric amplification and Higgs inflation} 	
The primordial curvature perturbations $\mathcal{R}$ of the  canonical inflation model in the Fourier space satisfy the Mukhanov-Sasaki equation~\cite{Kodama:1984ziu,Mukhanov:1990me}
 \begin{equation}\label{MSequation}
u''_{k}+(k^{2}-z''/z)u_{k}=0,
 \end{equation}
where $u=-z \mathcal{R}$, $z=a\dot{\phi}/H$, and 
\begin{equation}
    \frac{z''}{z}=2 a^{2} H^{2} \left( 1+\frac{3}{2}\delta +\epsilon +\frac{1}{2}\delta^{2} +\frac{1}{2}\epsilon\delta +\frac{1}{2H}\dot{\epsilon} +\frac{1}{2H}\dot{\delta} \right),
\end{equation}
with the slow-roll parameters $\epsilon \equiv -\dot{H}/H^{2}$ and $\delta \equiv \ddot{\phi}/H\dot{\phi}$ with the dot denoting the derivative with respect to the cosmic time $t$. Here, $H=\dot{a}/a$ is the Hubble parameter, and $a$ is the cosmic scale factor. To realize the parametric amplification mechanism, we assume that the potential consists of a small periodic structure $\delta V$, with $\delta V \ll \bar{V}$, added to a single-field slow-roll inflationary potential $\bar{V}$~\cite{Cai:2019bmk}. Specifically, the potential can be expressed as 
\begin{equation}\label{osc:potential}
        V(\phi)=\bar{V}(\phi)+\delta V(\phi)   
\end{equation}
where the oscillatory part is 
\begin{equation}\label{osc:part}
    \delta V(\phi)=\xi_0 \cos (\phi/\phi_{*}) \Theta(\phi;\phi_{s},\phi_{e}).
\end{equation}
Here,  $\xi_0, \phi_{*}, \phi_{s}$, and $\phi_{e}$ represent the magnitude, period, starting and ending points of the structure, respectively, and $\Theta(\phi;\phi_{s},\phi_{e})\equiv \Theta(\phi_{e}-\phi)\Theta(\phi-\phi_{s})$ is constructed by the Heaviside step function $\Theta$. We suppose that the small periodic structure has a negligible impact on the background evolution.  
During oscillatory phase (i.e.,  the time interval from $\phi_{s}$ to $\phi_{e}$),  $\dot{\phi}$ and  $\bar{V}_{,\phi}$ can be ignored, while $\ddot{\phi}$ and $\delta V_{,\phi}$ should be emphasized. Using the background equation,  we obtain the approximations $\phi \approx \phi_{s} + \dot{\phi}_{s}(t-t_{s})$ and $\ddot{\phi} \approx \xi_0 \sin(\phi/\phi_{*})/\phi_{*}$. Therefore, during the oscillatory phase, we have
$z''/z \approx a^{2} H \dot{\delta} \approx a^{2} \xi_0 \cos (\frac{\phi}{\phi_{*}})/\phi_{*}^{2}$,
if $H \ll 1$. 
This leads to the equation of motion for the perturbations $u_{k}$
\begin{equation}\label{eq:uk}
    \ddot{u}_{k} + H \dot{u}_{k} + \left[ \frac{k^{2}}{a^{2}} - \frac{\xi_0}{\phi_{*}^{2}}\cos(\frac{\phi}{\phi_{*}}) \right] u_{k} = 0.
\end{equation}

When focusing on the small scales ($k \gg a H$), the term $H \dot{u}_{k}$ is negligible. Consequently, equation~\eqref{eq:uk} can be rewritten as
\begin{equation}\label{eq:Mathieu}
    \frac{d^{2}u_{k}}{dx^{2}}+ \left[ A_{k}(x)-2q\cos 2x \right] u_{k} =0,
\end{equation}
where
$x= t \dot{\phi}_{s}/(2\phi_{*})+ (\phi_{s}-\dot{\phi}_{s}t_{s})/(2\phi_{*})$, $A_{k}(x)=k^{2}/(k_{*}^{2} a^{2})$, $k_{*}=|\dot{\phi}_{s}|/(2\phi_{*})$, and $q=2\xi_0/\dot{\phi}_{s}^{2}$. The  results of the power spectrum is
\begin{equation}\label{eq:pr}
    \mathcal{P}_{\mathcal{\mathcal{R}}}(k) \approx  A_{s} \left( \frac{k}{k_{p}} \right)^{n_{s}-1} \mathcal{A}^{2}(k).
\end{equation}
Here, $A_{s}$ and $n_{s}$ represent the amplitude and scalar spectral index of the power spectrum at the pivot scale $k_{p} = 0.05~\mathrm{Mpc}^{-1}$. The amplification factor $\mathcal{A}(k)$ is explicitly defined as 
\begin{equation}\label{eq:Ak}
\mathcal{A}(k) \equiv \frac{|u_{k}(t_{e})|}{|u_{k}(t_{s})|} \approx \exp \left( \int_{t_{s}}^{t_{e}} \mu_{k}(t) k_{*} dt \right),
\end{equation}
where
\begin{equation}\label{eq:mu}
    \mu_{k}(t) =\Re \left(  \sqrt{\left( \frac{q}{2} \right)^{2}-\left( \frac{k}{k_{*}a(t)}-1 \right)^{2}}  \right).
\end{equation}
Here, $\Re(y)$ denotes the real component of the quantity $y$.  To obtain the primordial curvature power spectrum, we also require the background evolution of $q(t)$ and $a(t)$ in equation \eqref{eq:mu}, which can only be obtained through numerical methods. 

In the following, we will consider the Higgs inflation model with non-minimal coupling term $\xi h^2 R/2$ and a small oscillatory part in the Higgs field.  
By employing the conformal transformation, the non-canonical Higgs inflation in the Jordan frame can be transformed into the canonical form in the Einstein frame. If there exists a small oscillatory part in the Higgs field, the potential in the Einstein frame will exhibit a similar oscillatory part, similar to equation \eqref{osc:potential}, thereby allowing for the application of parametric amplification.  
However, due to the high sensitivity of the parametric amplification mechanism to parameter values and the presence of the only approximate, rather than exact, relations between the two frames, exact relations like equations \eqref{eq:pr}-\eqref{eq:mu} cannot be derived. Despite this limitation, the primordial curvature power spectrum can indeed be enhanced through the parametric amplification mechanism, as similar equations to equations \eqref{eq:pr}-\eqref{eq:mu} exist in our model to enhance the primordial curvature power spectrum. To obtain the precise primordial curvature power spectrum, we numerically solve the background equations and the Mukhanov-Sasaki equation in our model.

For the Higgs inflation, the action is 
\cite{Bezrukov:2007ep}
\begin{equation}\label{act:higgs0}
S=\int d^4x \sqrt{-g} \left[\frac{M_{\mathrm{pl}}^2}{2} R +\xi H_\mathrm{SM}^\dag H_\mathrm{SM} R  +\mathcal{L}_\mathrm{SM}\right],
\end{equation}
where $\mathcal{L}_\mathrm{SM}$ is the Standard Model part, $H_\mathrm{SM}$ is the Higgs field, $\xi$ is a  dimensionless coupling constant, and $M_\mathrm{pl} = 1/\sqrt{8\pi G}$ is the reduced Planck mass. 
In the unitary gauge where $H_\mathrm{SM} = (0, h/\sqrt{2})^T$, the action of the Higgs inflation in the Jordan frame becomes 
\begin{equation}\label{act:higgs1}
    S= \int d^4x \sqrt{-g} \left[\frac{ M_{\mathrm{pl}}^2 +\xi h^2}{2}R-\frac{1}{2}(\partial h)^2 -V_h(h)\right].
\end{equation}
By using the condition that the energy of the Higgs field $h$ during the inflation epoch is significantly greater than the vacuum expectation value $v_\mathrm{EW}\simeq 250\, \mathrm{GeV}$, the Higgs potential can be expressed in a simplified form as $V_h(h) = \lambda h^4/4$.   In the following, we set the reduced Planck mass and the speed of light to unity, $ M_{\mathrm{pl}}=c=1$.  

For the homogeneous and isotropic Universe, the background equations are 
\begin{equation}\label{bgeq1}
  H^2+\frac{2  \xi  h  \dot{h}}{1+\xi  h ^2} H = \frac{1}{3 \left(1+\xi  h ^2\right)}\left[\frac{1}{2} \dot{h}^2+V_h(h)\right],
\end{equation}
\begin{equation}\label{bgeq2} 
A(h) (\ddot{h}+3H\dot{h})+B(h) \dot{h}^2+V_h'(h)=\frac{4\xi \phi V_h(h)}{1+\xi h^2},
\end{equation}
where $A(h) = 1+ 6 \xi ^2 h ^2 / (1+\xi  h^2)$, $B(h) = (\xi+6\xi^2)h / (1+\xi h^2)$, and $V_h'= dV_h/dh$.  
The equation for the primordial  curvature perturbation $\mathcal{R}_k$ satisfies the Mukhanov-Sasaki equation~\eqref{MSequation} with \cite{Hwang:1996bc}
\begin{equation}
    z = \frac{a \dot{h}}{H}\sqrt{\frac{\left(1 + \xi  h ^2\right) \left(1 + \xi h^2 +  6 \xi^2 h^2\right)}{\left( 1 + \xi    h ^2+ \xi  h  \dot{h}/H\right)^2}}.
\end{equation}

On the other hand, if the coupling constant $\xi$ is large enough, in the Einstein frame, the action for the Higgs inflation becomes  
\begin{equation}\label{act:higgs:e}
 S= \int d^4x \sqrt{-g} \left[\frac{1}{2}R-\frac{1}{2}(\partial\phi)^2 -V_{E}(\phi)  \right],
\end{equation}
where $\phi$ is the inflation field in the Einstein frame, and it can be related to the 
the inflation field $h$ in the Jordan frame  by  \cite{Bezrukov:2007ep}
\begin{equation}\label{field:chi}
   h =\frac{1}{\sqrt{\xi}}\exp\left(\frac{\phi}{\sqrt{6}}\right).
\end{equation} 
Note that the potential in the Einstein frame is  \cite{Bezrukov:2007ep}
\begin{equation}\label{einstein:V}
    V_E(\phi) \approx  \frac{\lambda}{4\xi^2}\left[1+\exp\left(-\frac{2\phi}{\sqrt{6}}\right)\right]^{-2}. 
\end{equation} 

To achieve the parametric amplification, the potential~\eqref{einstein:V} needs an extra small oscillatory part as displayed in Eqs.~\eqref{osc:potential} and \eqref{osc:part}. 
Combining  the approximate relation of the fields between the Jordan and Einstein frames \eqref{field:chi} and the cosine form \eqref{osc:part}, we can obtain a similar oscillatory part potential in the Jordan frame,  where the simplest form is 
\begin{equation}\label{higgs:delta}
   \delta V_h(h) =b \cos\left(\frac{1}{h_*}\ln h\right) \Theta(h;h_s,h_e).
\end{equation} 
Therefore, if the Higgs potential has a small oscillatory part $V(h)= V_h(h) +\delta V_h(h)$, the Mukhanov-Sasaki equation can become the Mathieu equation during the oscillatory phase, and the parametric amplification mechanism can be applied. Consequently, the primordial curvature power spectrum from the Higgs inflation can be enhanced.  
Furthermore, the oscillatory part \eqref{higgs:delta} only exists in the inflation region $h_e\leq h\leq h_s$, and disappears after the scale $h_e$, leading to the recovery of the usual Higgs inflation model.

%%%%%%%%%%%%%%%%%%%%%%%%%%%%%%%%%%%%%%%%%%%%%%%%%%%%%%%%%%%%%%%%%
\section{SIGWs and PBHs} 
Within the context of the Newtonian gauge, the perturbed metric is
\begin{equation}
ds^2 = a^2 \left\{-(1+2\Phi)\mathrm{d}\tau^2+[(1-2\Phi)\delta_{ij}+h_{ij}]\mathrm{d}x^i \mathrm{d}x^j\right\},
\end{equation}
where $\Phi$ denotes the Bardeen potential, and $h_{ij}$ are the tensor perturbations. We have ignored the influences from first-order gravitational waves, vector perturbations, and anisotropic stress as they are subdominant~\cite{Baumann:2007zm,Weinberg:2003ur,Watanabe:2006qe}. Following the Refs. \cite{Kohri:2018awv,Espinosa:2018eve}, the energy density of SIGWs at the epoch of matter-radiation equality can be presented as
\begin{equation}
\Omega_{\mathrm{GW}}(k) = \int_0^{\infty} \mathrm{d} v \int_{|1-v|}^{1+v} \mathrm{d} u \mathcal{T}(u, v) {P}_{\zeta}(v k) {P}_{\zeta}(u k),
\end{equation}
where ${P}_{\zeta}$ is the primordial power spectrum of curvature perturbations and 
 \begin{equation}
\begin{aligned}
\mathcal{T}(u,v)= & \frac{3}{1024 v^8 u^8}\left[4 v^2-\left(v^2-u^2+1\right)^2\right]^2\left(v^2+u^2-3\right)^2 \\
& \times\bigg\{\left[\left(v^2+u^2-3\right) \ln \left(\left|\frac{3-(v+u)^2}{3-(v-u)^2}\right|\right)-4 v u\right]^2 \\
& +\pi^2\left(v^2+u^2-3\right)^2 \Theta(v+u-\sqrt{3})\bigg\}.
\end{aligned}
\end{equation}
By utilizing the connection between the wave number $k$ and frequency $f$ that $k = 2\pi f$, one can express the energy density fraction spectrum of SIGWs at the current epoch as
\begin{equation}\label{omega:gw}
\Omega_{\mathrm{GW}, 0}(f)=\Omega_{\mathrm{r}, 0}\left[\frac{g_{, r}(T)}{g_{, r}\left(T_{\mathrm{eq}}\right)}\right]\left[\frac{g_{, s}\left(T_{\mathrm{eq}}\right)}{g_{, s}(T)}\right]^{\frac{4}{3}} \Omega_{\mathrm{GW}}(k).
\end{equation}
Here, $g_{,s}$ and $g_{,r}$ represent the effective degrees of freedom for entropy and radiation, respectively. Additionally, $\Omega_{r,0}$ is the present energy density fraction from radiation.

PBHs emerge through gravitational collapse, resulting from the density contrast $\delta\rho/\rho$ surpassing a critical threshold denoted as $\delta_c$ within Hubble patches. The connection between the PBH mass $M$ and wavenumber $k$ can be expressed as \cite{Inomata:2018cht}
\begin{equation}
\begin{aligned}
k &\approx \frac{1.6 \times 10^6}{\mathrm{Mpc}} \left(\frac{M_{\odot}}{M}\right)^{1 / 2}\left(\frac{\gamma}{0.2}\right)^{1/2}
 \left(\frac{g_{, r}\left(T\left(M\right)\right)}{106.75}\right)^{1 / 4}\left(\frac{g_{, s}\left(T\left(M\right)\right)}{106.75}\right)^{-1 / 3},
\end{aligned}
\end{equation}
where $M_{\odot}$ is the solar mass, and $\gamma\approx0.2$ represents the fraction of matter within the Hubble horizon that experiences gravitational collapse, consequently giving rise to the formation of PBHs. Here, we choose $\delta_c=0.42$. 
% The evaluation of the PBH abundance with their mass $M$, conventionally encompasses the characterization of $\beta(M)$ as the proportion of PBH masses with respect to the overall energy density during their formation epoch. 
The PBH abundance at formation can be formulated as a result of integrating the Gaussian distribution of perturbations, yielding
\begin{equation}
    \beta(M)=\frac{\gamma}{2}\text{erfc}\left(\frac{\delta_c}{\sqrt2\sigma(M)}\right).
\end{equation}
The parameter $\sigma(M)$, representing the variance of the density perturbation smoothed over the mass scale of $M$, is evaluated as 
\begin{equation}
\label{sigma}
\sigma^{2} =\frac{16}{81} ~ \int_{0}^{\infty}\! \frac{\mathrm{d}q}{q}\, \left(\frac{q}{k}\right)^{4} \tilde{W}^{2}(q/k) T^{2}(q/k) \mathcal{P}_{\mathcal{R}}(q).
\end{equation}
Here, $\tilde{W}(q/k)=\exp(-q^{2}/2k^{2})$ represents the Gaussian window function, while $T(q/k)=3(\sin l - l \cos l)/l^{3}$ stands for the transfer function, where $l = q/(\sqrt{3}k)$. The  total abundance of PBHs in the dark matter at present can be expressed as \cite{Sasaki:2018dmp}
\begin{equation}
f_{\mathrm{PBH}}\equiv \frac{\Omega_{\mathrm{PBH}}}{\Omega_{\mathrm{CDM}}} = \int f(M)\, \mathrm{d} \ln M,
\end{equation}
where $\Omega_{\mathrm{CDM}}$ is the density of cold dark matter and 
\begin{equation}\label{fM}
f(M) \approx 1.5 \times 10^8 \left(\frac{g_{,r}}{106.75}\right)^{3/4} \left(\frac{g_{,s}}{106.75}\right)^{-1} \left(\frac{M}{M_{\odot}}\right)^{-1/2} \beta(M).
\end{equation}

%%%%%%%%%%%%%%%%%%%%%%%%%%%%%%%%%%%%%%%%%%%%%%%%%%%%%%%%%%%%%%%%%
\section{Results} 
\begin{figure}[htbp]
    \centering
    \includegraphics[width=\linewidth]{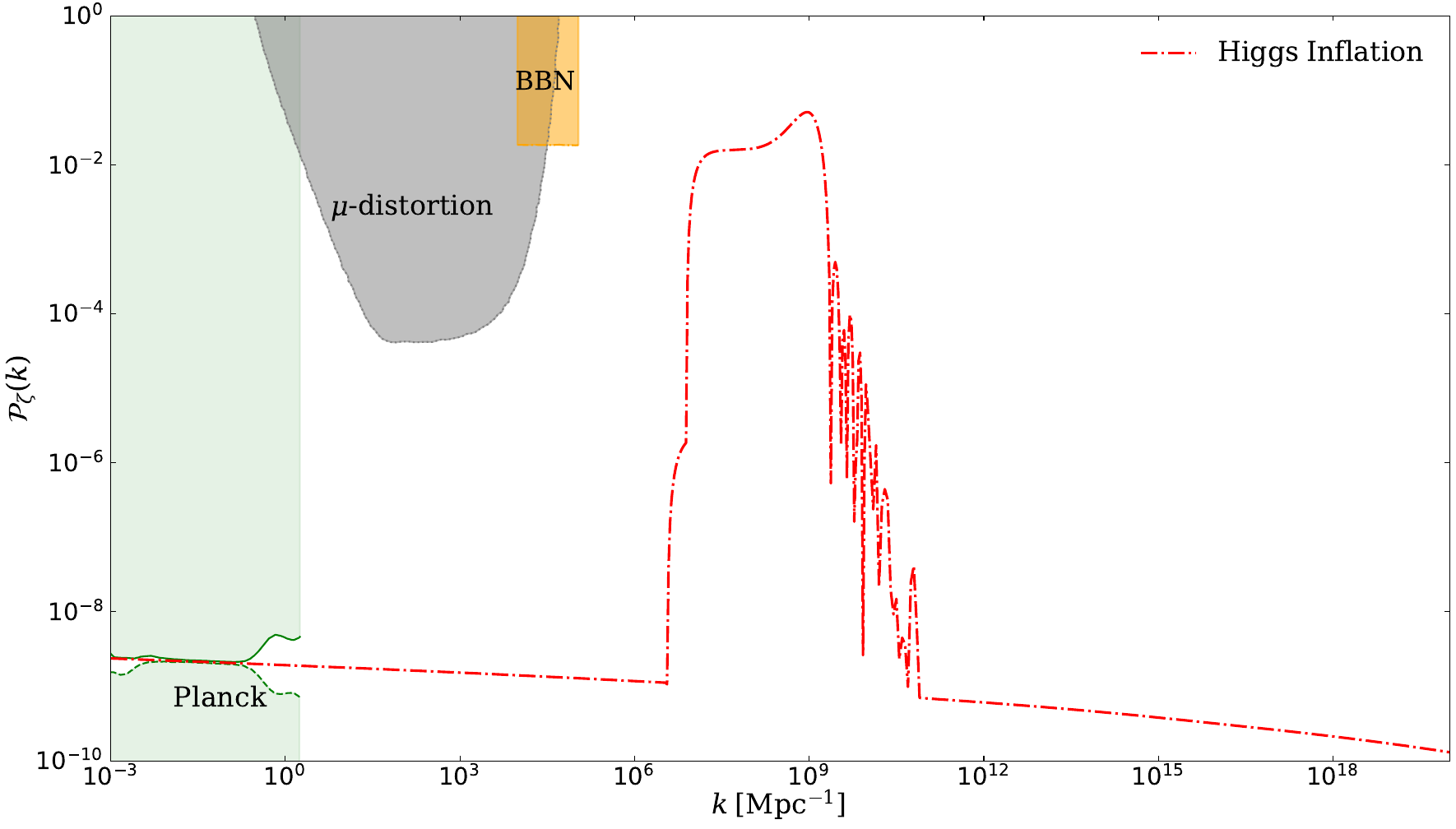}
    \caption{The power spectrum of curvature perturbations originates from the periodic structure of the Higgs inflation. The region shaded in light green is invalidated based on observations of the CMB \cite{Planck:2018jri}. The orange and gray regions denote constraints originating from the impact on the neutron-to-proton ratio during primordial nucleosynthesis (BBN)  \cite{Inomata:2016uip, Jeong:2014gna}, and the $\mu$-distortion of the CMB \cite{Fixsen:1996nj,Chluba:2012we}, respectively.}
    \label{fig:pr}
\end{figure}

\begin{figure}[htbp]
    \centering
    \includegraphics[width=\linewidth]{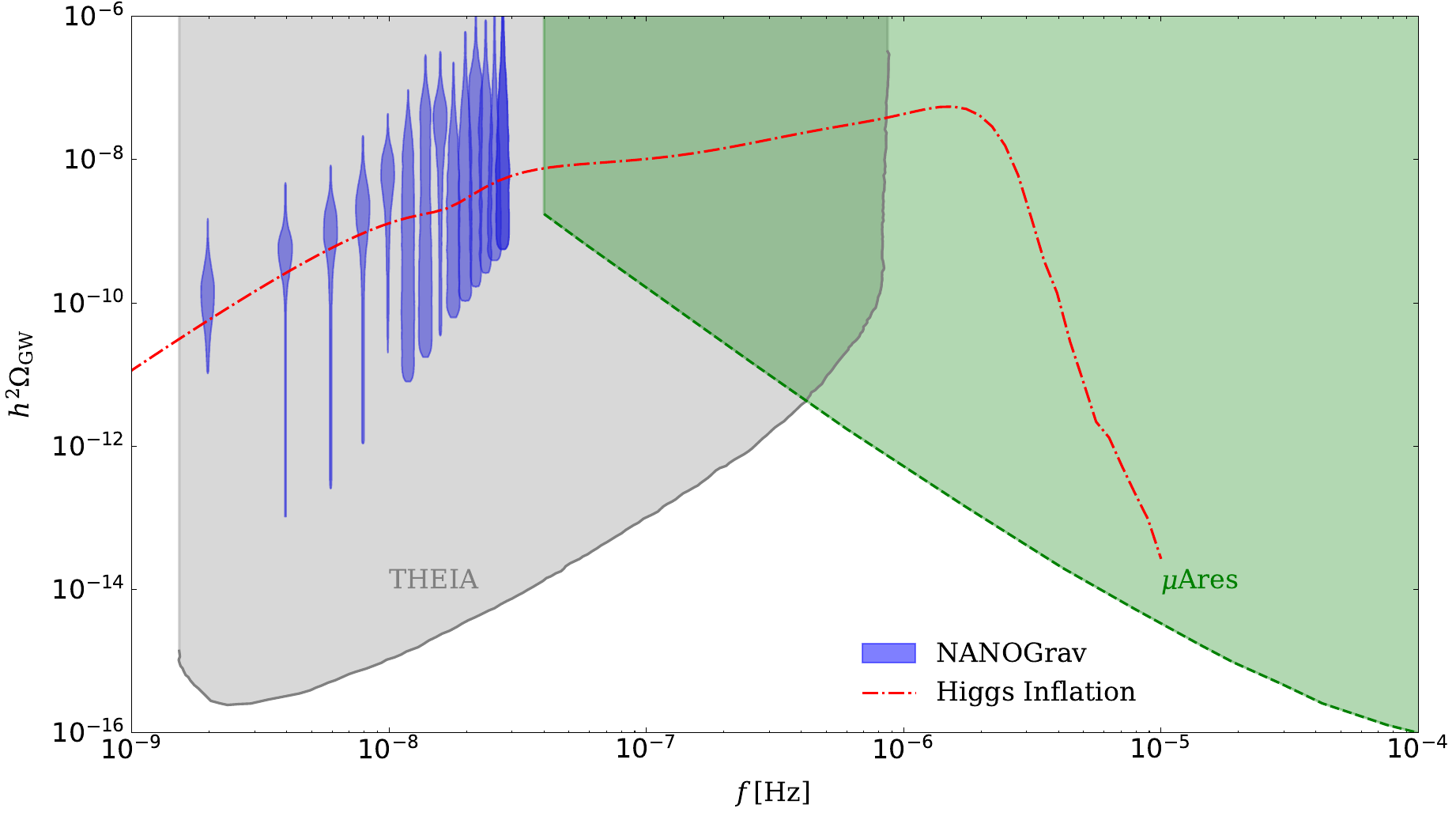}
    \caption{The red line is the energy density of SIGWs from the primordial curvature power spectrum displayed in Figure \ref{fig:pr}.  The blue violins are the free spectrum constraints from the NANOGrav 15-yr data set \cite{NANOGrav:2023gor, NANOGrav:2023hvm}. The gray and  green shaded areas are sensitive regions by future observation experiments of THEIA~\cite{Theia:2017xtk} and µAres~\cite{Sesana:2019vho}.}
    \label{fig:sigw}
\end{figure}

The parametric amplification mechanism can be realized in the Higgs inflation model if there exists a small extra oscillatory part \eqref{higgs:delta} in the Higgs field.  
To explore the NANOGrav signal and planet-mass PBHs,  we take the parameters  of our model as $\xi = 100$, $\lambda =3.37\times 10^{-6}$, $b =1.1\times10^{-11}$, $h_* = 5.89\times 10^{-6}$, $h_s=0.761$, and $h_e =0.754$, to  numerically solve the background equations \eqref{bgeq1} and \eqref{bgeq2}, and the Mukhanov-Sasaki equation. 
By using the definition of the primordial curvature power spectrum, 
\begin{equation}
\mathcal{P}_\mathcal{R} = \frac{k^3}{2\pi^2}|\mathcal{R}_k|^2,
\end{equation}
we can obtain the numerical result for the primordial curvature power spectrum as shown in Figure \ref{fig:pr}.  The amplitude of the primordial curvature power spectrum is amplified sharply. 
The inflation field at the horizon crossing of the pivot scale $k_p= 0.05 \, \mathrm{Mpc}^{-1}$ is $h_\mathrm{pivot}= 9.4/\sqrt{\xi}$ \cite{Bezrukov:2007ep} and the $e$-folds are $N=64$. The scalar spectral index is  $n_s = 0.969$, and the tensor-to-scalar ratio is $r= 0.0026$, consistent with the CMB observations~\cite{Planck:2018jri}.   
By using the numerical results of the primordial curvature power spectrum from the equation \eqref{omega:gw}, we can obtain the energy density of the corresponding SIGWs, which is displayed in Figure \ref{fig:sigw} and denoted by the red line. The blue violins are the free spectrum constraints from the NANOGrav 15-yr data set. It is shown that the energy density of the SIGWs from the Higgs inflation models can explain the NANOGrav 15-yr data set. 

\begin{figure}[htbp]
    \centering
    \includegraphics[width=\linewidth]{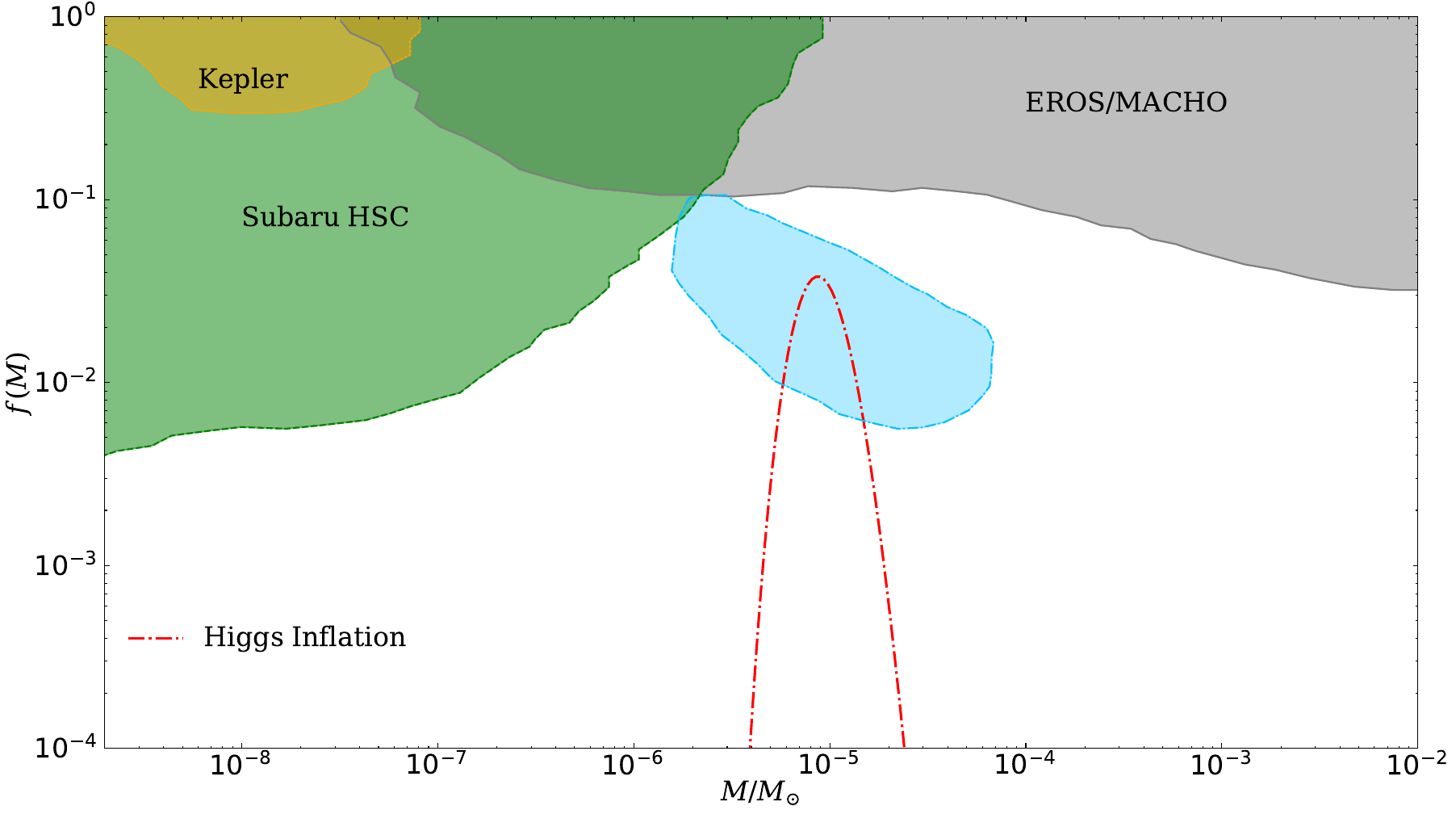}
    \caption{The red line shows the PBH abundance from the primordial curvature power spectrum displayed in Figure \ref{fig:pr}. The  shaded regions denote  the main observational constraints on the PBH abundance: 
		the gray region from the EROS/MACHO \cite{EROS-2:2006ryy}, the green region from microlensing events with Subaru HSC \cite{Niikura:2017zjd}, and the orange region from the Kepler satellite~\cite{Griest:2013esa}. The blue contour corresponds to the 95\% confidence level of the PBH abundance, assuming that the six lensing objects of planetary mass identified in the OGLE~\cite{Mroz:2017mvf} are PBHs~\cite{Niikura:2019kqi}.
    }
    \label{fig:fpbh}
\end{figure}

By using equation~\eqref{fM} and the numerical results of the primordial curvature power spectrum displayed in Figure~\ref{fig:pr}, we can get the abundance of the PBH, and the results are displayed in Figure \ref{fig:fpbh}, represented by the red line. {The abundance of PBHs is estimated to be $f_\text{PBH} = 0.024$, which is consistent with the constraint that the PBH abundance cannot exceed that of dark matter, i.e., $f_\text{PBH}\leq1$.} An interesting fact is that the corresponding abundance of the PBH is consistent with the implications of the ultrashort-timescale microlensing events recently observed by OGLE~\cite{Mroz:2017mvf,Niikura:2019kqi}, as shown in Figure \ref{fig:fpbh}.

In the above study, we have derived a set of parameters to explain both the NANOGrav 15-yr data set and the OGLE data. It is important to consider the sensitivity of the fitting  to changes in the model parameters. The parameter $\lambda$ is determined by the amplitude of the primordial curvature power spectrum at CMB scales and has little influence on the SIGW, so we do not investigate it further. In the enhancement region, the primordial curvature power spectrum is governed by the Mathieu equation \eqref{eq:Mathieu}. 
Parameter $h_*$ determines the oscillation speed of the Mathieu equation, while  $h_s$ and $h_e$ determine  the number of oscillation periods.  The parameter $h_e$  also determines the starting point of the enhancement region in the primordial curvature power spectrum.   
To explore the sensitivity of the fitting to variations in parameters, we employ a numerical approach, adjusting one parameter at a time while keeping the others constant, to obtain the likelihood function for explaining the NANOGrav 15-yr data set.  The likelihood function was obtained by evaluating the energy density of SIGWs at the 14 specific frequency bins of the NANOGrav 15-yr data set and then computing the product of the probability density functions from 14 independent kernel density estimates, which can be written as \cite{Moore:2021ibq}  
\begin{equation}
    \mathcal{L}(\Theta) = \prod_{i=1}^{14}\mathcal{L}_i(\Omega_{GW}(f_i,\Theta)),
\end{equation}
where $\Theta$ is one of the parameters $b$, $h_*$, $h_e$, or $h_e$.  
To characterize the  sensitivity, we use the  relative variation ratio 
\begin{equation}
    S_\Theta = \frac{|\Theta_e-\Theta_i|}{\Theta_0},
\end{equation} 
where $\Theta_i$ and $\Theta_e$ are the upper and lower limits of the parameter $\Theta$ giving  a non-zero likelihood,  and $\Theta_0$ is the value giving the maximum likelihood.  For the different parameters, the  relative variation ratios are  $S_b \approx 0.24$, $S_{h_*} \approx 1.1\times 10^{-5}$, $S_{h_e} \approx 1.32 \times 10^{-5}$, and $S_{h_s} = 2.6 \times 10^{-6}$.  The fitting is very sensitive to these parameters{, which can be understood as follows. Combining the oscillatory part of the potential \eqref{higgs:delta} in the Einstein frame with the Mathieu equation \eqref{eq:Mathieu}, we obtain 
\begin{equation}
q= \frac{2b}{[1+\xi h^2(\phi)]^2} \frac{1}{\dot{\phi}_s^2}, \quad x = \frac{\dot{\phi}_s t}{2\sqrt{6} h_*} - x_0,
\end{equation}
where $x_0=- (\phi_s-\sqrt{3/2}\ln \xi-\dot{\phi}_s t_s)/(2\sqrt{6}h_*)$. According to Floquet's theorem and disregarding the enhancement of the amplitude, the solution of Mathieu's equation \eqref{eq:Mathieu} has a period of either $\pi$ or $2\pi$. By using the relation \eqref{field:chi} and $\dot{\phi}_s \Delta t =\Delta \phi$, due to $h_* \sim 10^{-6}$, a shift of about $|\Delta h| \sim 10^{-6}$ in $h$ can lead to around one periodic shift in the solution of the Mathieu's equation, potentially resulting in a significant change in the amplitude of the scalar power spectrum. The limited range of observational data necessitates that the range of the solution of Mathieu's equation spans no more than one period.  Therefore, the sensitivities of the fitting to the parameters $h_e$ and $h_s$ are $S_{h_{e,s}} \lesssim 10^{-5}$. 
Regarding the parameter $h_*$, a relative shift of about $|\Delta h_*|/h_*\sim 10^{-4}$ will result in about one period change of the total number of periods as the solution evolves from $h_s$ to $h_e$. Hence, the sensitivity of the fitting to the parameter $h_*$ is $S_{h_{*}} \lesssim 10^{-4}$. As for the parameter $b$, it does not influence the periodicity of the solution but only affects its amplitude. While we cannot derive an effective formula to analyze its sensitivity, the linear relation of $q$ and $b$ may indicate that the amplitude of $u_k$ in equation \eqref{eq:Mathieu} is not very sensitive to the parameter $b$.
}

%%%%%%%%%%%%%%%%%%%%%%%%%%%%%%%%%%%%%%%%%%%%%%%%%%%%%%%%%%%%%%%%%
\section{Summary and discussion}
While the measurements obtained from the CMB and observations of large-scale structures have significantly improved our understanding of the Universe on a macroscopic level, our comprehension of smaller scales remains restricted, with the exception of the limits imposed by PBHs. Conversely, PTAs serve as an indispensable instrument for investigating the state of the early Universe by means of SIGWs. The recently published data from the NANOGrav presents persuasive evidence in favour of the presence of a stochastic signal that corresponds to an SGWB. Our research demonstrates that this signal can be attributed to the SIGWs originating from the Higgs inflation model, specifically through the parametric amplification mechanism. Such an SGWB naturally predicts the substantial existence of planet-mass PBHs, which could potentially account for Planet 9 within our solar system \cite{Scholtz:2019csj}, as well as serve as the lensing objects responsible for the ultrashort-timescale microlensing events observed by the OGLE \cite{Mroz:2017mvf,Niikura:2019kqi}. 
% The future observations of the SGWB through the utilization of PTAs and the study of planet-mass PBHs present an invaluable opportunity to probe Higgs inflation, which connects two fundamental aspects of theoretical physics: particle physics and cosmology. This framework could establish a bridge between the microscopic world of particle interactions and the macroscopic evolution of the Universe. 

The proposition that the nanoHerz SGWB and the planet-mass PBHs share the same origin is highly captivating, warranting further investigation through experiments involving PTA, microlensing, and direct exploration for the existence of Planet 9. It is worth acknowledging that although the current observations on the SGWB and PBHs do not provide definitive evidence, future experiments hold the potential to elucidate both of these phenomena. The sensitivity of nanoHertz SGWB can be significantly enhanced by SKA~\cite{Janssen:2014dka} by 3 to 5 orders. The utilization of Subaru HSC for microlensing search in the Andromeda Galaxy proves highly effective in mitigating the presence of unbounded planets, which predominantly reside in the galactic disk~\cite{Niikura:2019kqi,Kusenko:2020pcg}. Moreover, the inclusion of a PBH with a mass of around $10^{-5}~M_\odot$ within the solar system, referred to as Planet 9, may induce orbital irregularities in trans-Neptunian objects~\cite{Scholtz:2019csj}. The investigation of its surrounding minihalo~\cite{Siraj:2020upy}, Hawking radiation~\cite{Arbey:2020urq}, or gravitational field~\cite{Witten:2020ifl} could offer direct means to examine this phenomenon. 
Therefore, these ongoing and future experiments will offer valuable insights to further scrutinize and constrain our proposed scenario in the near future.

% Therefore, these experiments will either verify or falsify the predictions made by our proposed scenario in the near future.

%%%%%%%%%%%%%%%%%%%%%%%%%%%%%%%%%%%%%%%%%%%%%%%%%%%%%%%%%%%%%%%%%
\section*{Acknowledgments}
ZY is supported by the National Natural Science Foundation of China under Grant No. 12205015 and the supporting fund for young researcher of Beijing Normal University under Grant No. 28719/310432102.
ZQY is supported by the National Natural Science Foundation of China under Grant No. 12305059.
ZCC is supported by the National Natural Science Foundation of China (Grant No. 12247176 and No. 12247112) and the innovative research group of Hunan Province under Grant No. 2024JJ1006.
LL is supported by the National Natural Science Foundation of China (Grant No. 12247112 and No. 12247176) and the China Postdoctoral Science Foundation Fellowship No. 2023M730300. 

\bibliographystyle{JHEP}
\bibliography{ref}

\providecommand{\href}[2]{#2}\begingroup\raggedright\begin{thebibliography}{100}

\bibitem{Abbott:2016blz}
{\scshape LIGO Scientific, Virgo} collaboration, \emph{{Observation of
  Gravitational Waves from a Binary Black Hole Merger}},
  \href{https://doi.org/10.1103/PhysRevLett.116.061102}{\emph{Phys. Rev. Lett.}
  {\bfseries 116} (2016) 061102}
  [\href{https://arxiv.org/abs/1602.03837}{{\ttfamily 1602.03837}}].

\bibitem{TheLIGOScientific:2017qsa}
{\scshape LIGO Scientific, Virgo} collaboration, \emph{{GW170817: Observation
  of Gravitational Waves from a Binary Neutron Star Inspiral}},
  \href{https://doi.org/10.1103/PhysRevLett.119.161101}{\emph{Phys. Rev. Lett.}
  {\bfseries 119} (2017) 161101}
  [\href{https://arxiv.org/abs/1710.05832}{{\ttfamily 1710.05832}}].

\bibitem{LIGOScientific:2018mvr}
{\scshape LIGO Scientific, Virgo} collaboration, \emph{{GWTC-1: A
  Gravitational-Wave Transient Catalog of Compact Binary Mergers Observed by
  LIGO and Virgo during the First and Second Observing Runs}},
  \href{https://doi.org/10.1103/PhysRevX.9.031040}{\emph{Phys. Rev. X}
  {\bfseries 9} (2019) 031040}
  [\href{https://arxiv.org/abs/1811.12907}{{\ttfamily 1811.12907}}].

\bibitem{LIGOScientific:2020ibl}
{\scshape LIGO Scientific, Virgo} collaboration, \emph{{GWTC-2: Compact Binary
  Coalescences Observed by LIGO and Virgo During the First Half of the Third
  Observing Run}},
  \href{https://doi.org/10.1103/PhysRevX.11.021053}{\emph{Phys. Rev. X}
  {\bfseries 11} (2021) 021053}
  [\href{https://arxiv.org/abs/2010.14527}{{\ttfamily 2010.14527}}].

\bibitem{LIGOScientific:2021usb}
{\scshape LIGO Scientific, VIRGO} collaboration, \emph{{GWTC-2.1: Deep extended
  catalog of compact binary coalescences observed by LIGO and Virgo during the
  first half of the third observing run}},
  \href{https://doi.org/10.1103/PhysRevD.109.022001}{\emph{Phys. Rev. D}
  {\bfseries 109} (2024) 022001}
  [\href{https://arxiv.org/abs/2108.01045}{{\ttfamily 2108.01045}}].

\bibitem{LIGOScientific:2021djp}
{\scshape KAGRA, VIRGO, LIGO Scientific} collaboration, \emph{{GWTC-3: Compact
  Binary Coalescences Observed by LIGO and Virgo during the Second Part of the
  Third Observing Run}},
  \href{https://doi.org/10.1103/PhysRevX.13.041039}{\emph{Phys. Rev. X}
  {\bfseries 13} (2023) 041039}
  [\href{https://arxiv.org/abs/2111.03606}{{\ttfamily 2111.03606}}].

\bibitem{NANOGrav:2023gor}
{\scshape NANOGrav} collaboration, \emph{{The NANOGrav 15 yr Data Set: Evidence
  for a Gravitational-wave Background}},
  \href{https://doi.org/10.3847/2041-8213/acdac6}{\emph{Astrophys. J. Lett.}
  {\bfseries 951} (2023) L8}
  [\href{https://arxiv.org/abs/2306.16213}{{\ttfamily 2306.16213}}].

\bibitem{NANOGrav:2023hde}
{\scshape NANOGrav} collaboration, \emph{{The NANOGrav 15 yr Data Set:
  Observations and Timing of 68 Millisecond Pulsars}},
  \href{https://doi.org/10.3847/2041-8213/acda9a}{\emph{Astrophys. J. Lett.}
  {\bfseries 951} (2023) L9}
  [\href{https://arxiv.org/abs/2306.16217}{{\ttfamily 2306.16217}}].

\bibitem{Zic:2023gta}
A.~Zic et~al., \emph{{The Parkes Pulsar Timing Array third data release}},
  \href{https://doi.org/10.1017/pasa.2023.36}{\emph{Publ. Astron. Soc.
  Austral.} {\bfseries 40} (2023) e049}
  [\href{https://arxiv.org/abs/2306.16230}{{\ttfamily 2306.16230}}].

\bibitem{Reardon:2023gzh}
D.J.~Reardon et~al., \emph{{Search for an Isotropic Gravitational-wave
  Background with the Parkes Pulsar Timing Array}},
  \href{https://doi.org/10.3847/2041-8213/acdd02}{\emph{Astrophys. J. Lett.}
  {\bfseries 951} (2023) L6}
  [\href{https://arxiv.org/abs/2306.16215}{{\ttfamily 2306.16215}}].

\bibitem{EPTA:2023sfo}
{\scshape EPTA} collaboration, \emph{{The second data release from the European
  Pulsar Timing Array - I. The dataset and timing analysis}},
  \href{https://doi.org/10.1051/0004-6361/202346841}{\emph{Astron. Astrophys.}
  {\bfseries 678} (2023) A48}
  [\href{https://arxiv.org/abs/2306.16224}{{\ttfamily 2306.16224}}].

\bibitem{Antoniadis:2023ott}
{\scshape EPTA, InPTA:} collaboration, \emph{{The second data release from the
  European Pulsar Timing Array - III. Search for gravitational wave signals}},
  \href{https://doi.org/10.1051/0004-6361/202346844}{\emph{Astron. Astrophys.}
  {\bfseries 678} (2023) A50}
  [\href{https://arxiv.org/abs/2306.16214}{{\ttfamily 2306.16214}}].

\bibitem{Xu:2023wog}
H.~Xu et~al., \emph{{Searching for the Nano-Hertz Stochastic Gravitational Wave
  Background with the Chinese Pulsar Timing Array Data Release I}},
  \href{https://doi.org/10.1088/1674-4527/acdfa5}{\emph{Res. Astron.
  Astrophys.} {\bfseries 23} (2023) 075024}
  [\href{https://arxiv.org/abs/2306.16216}{{\ttfamily 2306.16216}}].

\bibitem{Hellings:1983fr}
R.w.~Hellings and G.s.~Downs, \emph{{UPPER LIMITS ON THE ISOTROPIC
  GRAVITATIONAL RADIATION BACKGROUND FROM PULSAR TIMING ANALYSIS}},
  \href{https://doi.org/10.1086/183954}{\emph{Astrophys. J. Lett.} {\bfseries
  265} (1983) L39}.

\bibitem{Li:2019vlb}
J.~Li, Z.-C.~Chen and Q.-G.~Huang, \emph{{Measuring the tilt of primordial
  gravitational-wave power spectrum from observations}},
  \href{https://doi.org/10.1007/s11433-019-9605-5}{\emph{Sci. China Phys. Mech.
  Astron.} {\bfseries 62} (2019) 110421}
  [\href{https://arxiv.org/abs/1907.09794}{{\ttfamily 1907.09794}}].

\bibitem{Vagnozzi:2020gtf}
S.~Vagnozzi, \emph{{Implications of the NANOGrav results for inflation}},
  \href{https://doi.org/10.1093/mnrasl/slaa203}{\emph{Mon. Not. Roy. Astron.
  Soc.} {\bfseries 502} (2021) L11}
  [\href{https://arxiv.org/abs/2009.13432}{{\ttfamily 2009.13432}}].

\bibitem{Chen:2021wdo}
Z.-C.~Chen, C.~Yuan and Q.-G.~Huang, \emph{{Non-tensorial gravitational wave
  background in NANOGrav 12.5-year data set}},
  \href{https://doi.org/10.1007/s11433-021-1797-y}{\emph{Sci. China Phys. Mech.
  Astron.} {\bfseries 64} (2021) 120412}
  [\href{https://arxiv.org/abs/2101.06869}{{\ttfamily 2101.06869}}].

\bibitem{Wu:2021kmd}
Y.-M.~Wu, Z.-C.~Chen and Q.-G.~Huang, \emph{{Constraining the Polarization of
  Gravitational Waves with the Parkes Pulsar Timing Array Second Data
  Release}}, \href{https://doi.org/10.3847/1538-4357/ac35cc}{\emph{Astrophys.
  J.} {\bfseries 925} (2022) 37}
  [\href{https://arxiv.org/abs/2108.10518}{{\ttfamily 2108.10518}}].

\bibitem{Chen:2021ncc}
Z.-C.~Chen, Y.-M.~Wu and Q.-G.~Huang, \emph{{Searching for isotropic stochastic
  gravitational-wave background in the international pulsar timing array second
  data release}}, \href{https://doi.org/10.1088/1572-9494/ac7cdf}{\emph{Commun.
  Theor. Phys.} {\bfseries 74} (2022) 105402}
  [\href{https://arxiv.org/abs/2109.00296}{{\ttfamily 2109.00296}}].

\bibitem{Sakharov:2021dim}
A.S.~Sakharov, Y.N.~Eroshenko and S.G.~Rubin, \emph{{Looking at the NANOGrav
  signal through the anthropic window of axionlike particles}},
  \href{https://doi.org/10.1103/PhysRevD.104.043005}{\emph{Phys. Rev. D}
  {\bfseries 104} (2021) 043005}
  [\href{https://arxiv.org/abs/2104.08750}{{\ttfamily 2104.08750}}].

\bibitem{Benetti:2021uea}
M.~Benetti, L.L.~Graef and S.~Vagnozzi, \emph{{Primordial gravitational waves
  from NANOGrav: A broken power-law approach}},
  \href{https://doi.org/10.1103/PhysRevD.105.043520}{\emph{Phys. Rev. D}
  {\bfseries 105} (2022) 043520}
  [\href{https://arxiv.org/abs/2111.04758}{{\ttfamily 2111.04758}}].

\bibitem{Chen:2022azo}
Z.-C.~Chen, Y.-M.~Wu and Q.-G.~Huang, \emph{{Search for the Gravitational-wave
  Background from Cosmic Strings with the Parkes Pulsar Timing Array Second
  Data Release}},
  \href{https://doi.org/10.3847/1538-4357/ac86cb}{\emph{Astrophys. J.}
  {\bfseries 936} (2022) 20}
  [\href{https://arxiv.org/abs/2205.07194}{{\ttfamily 2205.07194}}].

\bibitem{Ashoorioon:2022raz}
A.~Ashoorioon, K.~Rezazadeh and A.~Rostami, \emph{{NANOGrav signal from the end
  of inflation and the LIGO mass and heavier primordial black holes}},
  \href{https://doi.org/10.1016/j.physletb.2022.137542}{\emph{Phys. Lett. B}
  {\bfseries 835} (2022) 137542}
  [\href{https://arxiv.org/abs/2202.01131}{{\ttfamily 2202.01131}}].

\bibitem{PPTA:2022eul}
{\scshape PPTA} collaboration, \emph{{Constraining ultralight vector dark
  matter with the Parkes Pulsar Timing Array second data release}},
  \href{https://doi.org/10.1103/PhysRevD.106.L081101}{\emph{Phys. Rev. D}
  {\bfseries 106} (2022) L081101}
  [\href{https://arxiv.org/abs/2210.03880}{{\ttfamily 2210.03880}}].

\bibitem{Wu:2023pbt}
Y.-M.~Wu, Z.-C.~Chen and Q.-G.~Huang, \emph{{Search for stochastic
  gravitational-wave background from massive gravity in the NANOGrav 12.5-year
  dataset}}, \href{https://doi.org/10.1103/PhysRevD.107.042003}{\emph{Phys.
  Rev. D} {\bfseries 107} (2023) 042003}
  [\href{https://arxiv.org/abs/2302.00229}{{\ttfamily 2302.00229}}].

\bibitem{IPTA:2023ero}
{\scshape IPTA} collaboration, \emph{{Searching for continuous Gravitational
  Waves in the second data release of the International Pulsar Timing Array}},
  \href{https://doi.org/10.1093/mnras/stad812}{\emph{Mon. Not. Roy. Astron.
  Soc.} {\bfseries 521} (2023) 5077}
  [\href{https://arxiv.org/abs/2303.10767}{{\ttfamily 2303.10767}}].

\bibitem{Wu:2023dnp}
Y.-M.~Wu, Z.-C.~Chen and Q.-G.~Huang, \emph{{Pulsar timing residual induced by
  ultralight tensor dark matter}},
  \href{https://doi.org/10.1088/1475-7516/2023/09/021}{\emph{JCAP} {\bfseries
  09} (2023) 021} [\href{https://arxiv.org/abs/2305.08091}{{\ttfamily
  2305.08091}}].

\bibitem{Dandoy:2023jot}
V.~Dandoy, V.~Domcke and F.~Rompineve, \emph{{Search for scalar induced
  gravitational waves in the international pulsar timing array data release 2
  and NANOgrav 12.5 years datasets}},
  \href{https://doi.org/10.21468/SciPostPhysCore.6.3.060}{\emph{SciPost Phys.
  Core} {\bfseries 6} (2023) 060}
  [\href{https://arxiv.org/abs/2302.07901}{{\ttfamily 2302.07901}}].

\bibitem{Madge:2023cak}
E.~Madge, E.~Morgante, C.~Puchades-Ib\'a\~nez, N.~Ramberg, W.~Ratzinger,
  S.~Schenk et~al., \emph{{Primordial gravitational waves in the nano-Hertz
  regime and PTA data \textemdash{} towards solving the GW inverse problem}},
  \href{https://doi.org/10.1007/JHEP10(2023)171}{\emph{JHEP} {\bfseries 10}
  (2023) 171} [\href{https://arxiv.org/abs/2306.14856}{{\ttfamily
  2306.14856}}].

\bibitem{Chen:2023zkb}
Z.-C.~Chen, Q.-G.~Huang, C.~Liu, L.~Liu, X.-J.~Liu, Y.~Wu et~al.,
  \emph{{Prospects for Taiji to detect a gravitational-wave background from
  cosmic strings}},
  \href{https://doi.org/10.1088/1475-7516/2024/03/022}{\emph{JCAP} {\bfseries
  03} (2024) 022} [\href{https://arxiv.org/abs/2310.00411}{{\ttfamily
  2310.00411}}].

\bibitem{NANOGrav:2023hvm}
{\scshape NANOGrav} collaboration, \emph{{The NANOGrav 15 yr Data Set: Search
  for Signals from New Physics}},
  \href{https://doi.org/10.3847/2041-8213/acdc91}{\emph{Astrophys. J. Lett.}
  {\bfseries 951} (2023) L11}
  [\href{https://arxiv.org/abs/2306.16219}{{\ttfamily 2306.16219}}].

\bibitem{Antoniadis:2023xlr}
{\scshape EPTA} collaboration, \emph{{The second data release from the European
  Pulsar Timing Array: V. Implications for massive black holes, dark matter and
  the early Universe}},  \href{https://arxiv.org/abs/2306.16227}{{\ttfamily
  2306.16227}}.

\bibitem{King:2023cgv}
S.F.~King, D.~Marfatia and M.H.~Rahat, \emph{{Toward distinguishing Dirac from
  Majorana neutrino mass with gravitational waves}},
  \href{https://doi.org/10.1103/PhysRevD.109.035014}{\emph{Phys. Rev. D}
  {\bfseries 109} (2024) 035014}
  [\href{https://arxiv.org/abs/2306.05389}{{\ttfamily 2306.05389}}].

\bibitem{Niu:2023bsr}
X.~Niu and M.H.~Rahat, \emph{{NANOGrav signal from axion inflation}},
  \href{https://doi.org/10.1103/PhysRevD.108.115023}{\emph{Phys. Rev. D}
  {\bfseries 108} (2023) 115023}
  [\href{https://arxiv.org/abs/2307.01192}{{\ttfamily 2307.01192}}].

\bibitem{Bi:2023tib}
Y.-C.~Bi, Y.-M.~Wu, Z.-C.~Chen and Q.-G.~Huang, \emph{{Implications for the
  supermassive black hole binaries from the NANOGrav 15-year data set}},
  \href{https://doi.org/10.1007/s11433-023-2252-4}{\emph{Sci. China Phys. Mech.
  Astron.} {\bfseries 66} (2023) 120402}
  [\href{https://arxiv.org/abs/2307.00722}{{\ttfamily 2307.00722}}].

\bibitem{Liu:2023pau}
L.~Liu, Z.-C.~Chen and Q.-G.~Huang, \emph{{Probing the equation of state of the
  early Universe with pulsar timing arrays}},
  \href{https://doi.org/10.1088/1475-7516/2023/11/071}{\emph{JCAP} {\bfseries
  11} (2023) 071} [\href{https://arxiv.org/abs/2307.14911}{{\ttfamily
  2307.14911}}].

\bibitem{Vagnozzi:2023lwo}
S.~Vagnozzi, \emph{{Inflationary interpretation of the stochastic gravitational
  wave background signal detected by pulsar timing array experiments}},
  \href{https://doi.org/10.1016/j.jheap.2023.07.001}{\emph{JHEAp} {\bfseries
  39} (2023) 81} [\href{https://arxiv.org/abs/2306.16912}{{\ttfamily
  2306.16912}}].

\bibitem{Han:2023olf}
C.~Han, K.-P.~Xie, J.M.~Yang and M.~Zhang, \emph{{Self-interacting dark matter
  implied by nano-Hertz gravitational waves}},
  \href{https://arxiv.org/abs/2306.16966}{{\ttfamily 2306.16966}}.

\bibitem{Li:2023yaj}
Y.-Y.~Li, C.~Zhang, Z.~Wang, M.-Y.~Cui, Y.-L.S.~Tsai, Q.~Yuan et~al.,
  \emph{{Primordial magnetic field as a common solution of nanohertz
  gravitational waves and the Hubble tension}},
  \href{https://doi.org/10.1103/PhysRevD.109.043538}{\emph{Phys. Rev. D}
  {\bfseries 109} (2024) 043538}
  [\href{https://arxiv.org/abs/2306.17124}{{\ttfamily 2306.17124}}].

\bibitem{Franciolini:2023wjm}
G.~Franciolini, D.~Racco and F.~Rompineve, \emph{{Footprints of the QCD
  Crossover on Cosmological Gravitational Waves at Pulsar Timing Arrays}},
  \href{https://doi.org/10.1103/PhysRevLett.132.081001}{\emph{Phys. Rev. Lett.}
  {\bfseries 132} (2024) 081001}
  [\href{https://arxiv.org/abs/2306.17136}{{\ttfamily 2306.17136}}].

\bibitem{Shen:2023pan}
Z.-Q.~Shen, G.-W.~Yuan, Y.-Y.~Wang and Y.-Z.~Wang, \emph{{Dark Matter Spike
  surrounding Supermassive Black Holes Binary and the nanohertz Stochastic
  Gravitational Wave Background}},
  \href{https://arxiv.org/abs/2306.17143}{{\ttfamily 2306.17143}}.

\bibitem{Kitajima:2023cek}
N.~Kitajima, J.~Lee, K.~Murai, F.~Takahashi and W.~Yin, \emph{{Gravitational
  waves from domain wall collapse, and application to nanohertz signals with
  QCD-coupled axions}},
  \href{https://doi.org/10.1016/j.physletb.2024.138586}{\emph{Phys. Lett. B}
  {\bfseries 851} (2024) 138586}
  [\href{https://arxiv.org/abs/2306.17146}{{\ttfamily 2306.17146}}].

\bibitem{Franciolini:2023pbf}
G.~Franciolini, A.~Iovino, Junior., V.~Vaskonen and H.~Veermae, \emph{{Recent
  Gravitational Wave Observation by Pulsar Timing Arrays and Primordial Black
  Holes: The Importance of Non-Gaussianities}},
  \href{https://doi.org/10.1103/PhysRevLett.131.201401}{\emph{Phys. Rev. Lett.}
  {\bfseries 131} (2023) 201401}
  [\href{https://arxiv.org/abs/2306.17149}{{\ttfamily 2306.17149}}].

\bibitem{Addazi:2023jvg}
A.~Addazi, Y.-F.~Cai, A.~Marciano and L.~Visinelli, \emph{{Have pulsar timing
  array methods detected a cosmological phase transition?}},
  \href{https://doi.org/10.1103/PhysRevD.109.015028}{\emph{Phys. Rev. D}
  {\bfseries 109} (2024) 015028}
  [\href{https://arxiv.org/abs/2306.17205}{{\ttfamily 2306.17205}}].

\bibitem{Cai:2023dls}
Y.-F.~Cai, X.-C.~He, X.-H.~Ma, S.-F.~Yan and G.-W.~Yuan, \emph{{Limits on
  scalar-induced gravitational waves from the stochastic background by pulsar
  timing array observations}},
  \href{https://doi.org/10.1016/j.scib.2023.10.027}{\emph{Sci. Bull.}
  {\bfseries 68} (2023) 2929}
  [\href{https://arxiv.org/abs/2306.17822}{{\ttfamily 2306.17822}}].

\bibitem{Inomata:2023zup}
K.~Inomata, K.~Kohri and T.~Terada, \emph{{Detected stochastic gravitational
  waves and subsolar-mass primordial black holes}},
  \href{https://doi.org/10.1103/PhysRevD.109.063506}{\emph{Phys. Rev. D}
  {\bfseries 109} (2024) 063506}
  [\href{https://arxiv.org/abs/2306.17834}{{\ttfamily 2306.17834}}].

\bibitem{Murai:2023gkv}
K.~Murai and W.~Yin, \emph{{A novel probe of supersymmetry in light of
  nanohertz gravitational waves}},
  \href{https://doi.org/10.1007/JHEP10(2023)062}{\emph{JHEP} {\bfseries 10}
  (2023) 062} [\href{https://arxiv.org/abs/2307.00628}{{\ttfamily
  2307.00628}}].

\bibitem{Li:2023bxy}
S.-P.~Li and K.-P.~Xie, \emph{{Collider test of nano-Hertz gravitational waves
  from pulsar timing arrays}},
  \href{https://doi.org/10.1103/PhysRevD.108.055018}{\emph{Phys. Rev. D}
  {\bfseries 108} (2023) 055018}
  [\href{https://arxiv.org/abs/2307.01086}{{\ttfamily 2307.01086}}].

\bibitem{Anchordoqui:2023tln}
L.A.~Anchordoqui, I.~Antoniadis and D.~Lust, \emph{{Fuzzy dark matter and the
  dark dimension}},
  \href{https://doi.org/10.1140/epjc/s10052-024-12622-y}{\emph{Eur. Phys. J. C}
  {\bfseries 84} (2024) 273}
  [\href{https://arxiv.org/abs/2307.01100}{{\ttfamily 2307.01100}}].

\bibitem{Liu:2023ymk}
L.~Liu, Z.-C.~Chen and Q.-G.~Huang, \emph{{Implications for the non-Gaussianity
  of curvature perturbation from pulsar timing arrays}},
  \href{https://doi.org/10.1103/PhysRevD.109.L061301}{\emph{Phys. Rev. D}
  {\bfseries 109} (2024) L061301}
  [\href{https://arxiv.org/abs/2307.01102}{{\ttfamily 2307.01102}}].

\bibitem{Abe:2023yrw}
K.T.~Abe and Y.~Tada, \emph{{Translating nano-Hertz gravitational wave
  background into primordial perturbations taking account of the cosmological
  QCD phase transition}},
  \href{https://doi.org/10.1103/PhysRevD.108.L101304}{\emph{Phys. Rev. D}
  {\bfseries 108} (2023) L101304}
  [\href{https://arxiv.org/abs/2307.01653}{{\ttfamily 2307.01653}}].

\bibitem{Ghosh:2023aum}
T.~Ghosh, A.~Ghoshal, H.-K.~Guo, F.~Hajkarim, S.F.~King, K.~Sinha et~al.,
  \emph{{Did we hear the sound of the Universe boiling? Analysis using the full
  fluid velocity profiles and NANOGrav 15-year data}},
  \href{https://arxiv.org/abs/2307.02259}{{\ttfamily 2307.02259}}.

\bibitem{Figueroa:2023zhu}
D.G.~Figueroa, M.~Pieroni, A.~Ricciardone and P.~Simakachorn,
  \emph{{Cosmological Background Interpretation of Pulsar Timing Array Data}},
  \href{https://arxiv.org/abs/2307.02399}{{\ttfamily 2307.02399}}.

\bibitem{Yi:2023mbm}
Z.~Yi, Q.~Gao, Y.~Gong, Y.~Wang and F.~Zhang, \emph{{Scalar induced
  gravitational waves in light of Pulsar Timing Array data}},
  \href{https://doi.org/10.1007/s11433-023-2266-1}{\emph{Sci. China Phys. Mech.
  Astron.} {\bfseries 66} (2023) 120404}
  [\href{https://arxiv.org/abs/2307.02467}{{\ttfamily 2307.02467}}].

\bibitem{Wu:2023hsa}
Y.-M.~Wu, Z.-C.~Chen and Q.-G.~Huang, \emph{{Cosmological interpretation for
  the stochastic signal in pulsar timing arrays}},
  \href{https://doi.org/10.1007/s11433-023-2298-7}{\emph{Sci. China Phys. Mech.
  Astron.} {\bfseries 67} (2024) 240412}
  [\href{https://arxiv.org/abs/2307.03141}{{\ttfamily 2307.03141}}].

\bibitem{Li:2023tdx}
X.-F.~Li, \emph{{Probing the high temperature symmetry breaking with
  gravitational waves from domain walls}},
  \href{https://arxiv.org/abs/2307.03163}{{\ttfamily 2307.03163}}.

\bibitem{Geller:2023shn}
M.~Geller, S.~Ghosh, S.~Lu and Y.~Tsai, \emph{{Challenges in interpreting the
  NANOGrav 15-year dataset as early Universe gravitational waves produced by an
  ALP induced instability}},
  \href{https://doi.org/10.1103/PhysRevD.109.063537}{\emph{Phys. Rev. D}
  {\bfseries 109} (2024) 063537}
  [\href{https://arxiv.org/abs/2307.03724}{{\ttfamily 2307.03724}}].

\bibitem{You:2023rmn}
Z.-Q.~You, Z.~Yi and Y.~Wu, \emph{{Constraints on primordial curvature power
  spectrum with pulsar timing arrays}},
  \href{https://doi.org/10.1088/1475-7516/2023/11/065}{\emph{JCAP} {\bfseries
  11} (2023) 065} [\href{https://arxiv.org/abs/2307.04419}{{\ttfamily
  2307.04419}}].

\bibitem{Antusch:2023zjk}
S.~Antusch, K.~Hinze, S.~Saad and J.~Steiner, \emph{{Singling out SO(10) GUT
  models using recent PTA results}},
  \href{https://doi.org/10.1103/PhysRevD.108.095053}{\emph{Phys. Rev. D}
  {\bfseries 108} (2023) 095053}
  [\href{https://arxiv.org/abs/2307.04595}{{\ttfamily 2307.04595}}].

\bibitem{Ye:2023xyr}
G.~Ye and A.~Silvestri, \emph{{Can the Gravitational Wave Background Feel
  Wiggles in Spacetime?}},
  \href{https://doi.org/10.3847/2041-8213/ad2851}{\emph{Astrophys. J. Lett.}
  {\bfseries 963} (2024) L15}
  [\href{https://arxiv.org/abs/2307.05455}{{\ttfamily 2307.05455}}].

\bibitem{HosseiniMansoori:2023mqh}
S.A.~Hosseini~Mansoori, F.~Felegray, A.~Talebian and M.~Sami, \emph{{PBHs and
  GWs from \ensuremath{\mathbb{T}}$^{2}$-inflation and NANOGrav 15-year data}},
  \href{https://doi.org/10.1088/1475-7516/2023/08/067}{\emph{JCAP} {\bfseries
  08} (2023) 067} [\href{https://arxiv.org/abs/2307.06757}{{\ttfamily
  2307.06757}}].

\bibitem{Jin:2023wri}
J.-H.~Jin, Z.-C.~Chen, Z.~Yi, Z.-Q.~You, L.~Liu and Y.~Wu, \emph{{Confronting
  sound speed resonance with pulsar timing arrays}},
  \href{https://doi.org/10.1088/1475-7516/2023/09/016}{\emph{JCAP} {\bfseries
  09} (2023) 016} [\href{https://arxiv.org/abs/2307.08687}{{\ttfamily
  2307.08687}}].

\bibitem{Zhang:2023nrs}
Z.~Zhang, C.~Cai, Y.-H.~Su, S.~Wang, Z.-H.~Yu and H.-H.~Zhang,
  \emph{{Nano-Hertz gravitational waves from collapsing domain walls associated
  with freeze-in dark matter in light of pulsar timing array observations}},
  \href{https://doi.org/10.1103/PhysRevD.108.095037}{\emph{Phys. Rev. D}
  {\bfseries 108} (2023) 095037}
  [\href{https://arxiv.org/abs/2307.11495}{{\ttfamily 2307.11495}}].

\bibitem{ValbusaDallArmi:2023nqn}
L.~Valbusa~Dall'Armi, A.~Mierna, S.~Matarrese and A.~Ricciardone,
  \emph{{Adiabatic or Non-Adiabatic? Unraveling the Nature of Initial
  Conditions in the Cosmological Gravitational Wave Background}},
  \href{https://arxiv.org/abs/2307.11043}{{\ttfamily 2307.11043}}.

\bibitem{DeLuca:2023tun}
V.~De~Luca, A.~Kehagias and A.~Riotto, \emph{{How well do we know the
  primordial black hole abundance: The crucial role of nonlinearities when
  approaching the horizon}},
  \href{https://doi.org/10.1103/PhysRevD.108.063531}{\emph{Phys. Rev. D}
  {\bfseries 108} (2023) 063531}
  [\href{https://arxiv.org/abs/2307.13633}{{\ttfamily 2307.13633}}].

\bibitem{Choudhury:2023kam}
S.~Choudhury, \emph{{Single field inflation in the light of Pulsar Timing Array
  Data: quintessential interpretation of blue tilted tensor spectrum through
  Non-Bunch Davies initial condition}},
  \href{https://doi.org/10.1140/epjc/s10052-024-12625-9}{\emph{Eur. Phys. J. C}
  {\bfseries 84} (2024) 278}
  [\href{https://arxiv.org/abs/2307.03249}{{\ttfamily 2307.03249}}].

\bibitem{Gorji:2023sil}
M.A.~Gorji, M.~Sasaki and T.~Suyama, \emph{{Extra-tensor-induced origin for the
  PTA signal: No primordial black hole production}},
  \href{https://doi.org/10.1016/j.physletb.2023.138214}{\emph{Phys. Lett. B}
  {\bfseries 846} (2023) 138214}
  [\href{https://arxiv.org/abs/2307.13109}{{\ttfamily 2307.13109}}].

\bibitem{Das:2023nmm}
B.~Das, N.~Jaman and M.~Sami, \emph{{Gravitational wave background from
  quintessential inflation and NANOGrav data}},
  \href{https://doi.org/10.1103/PhysRevD.108.103510}{\emph{Phys. Rev. D}
  {\bfseries 108} (2023) 103510}
  [\href{https://arxiv.org/abs/2307.12913}{{\ttfamily 2307.12913}}].

\bibitem{Yi:2023tdk}
Z.~Yi, Z.-Q.~You and Y.~Wu, \emph{{Model-independent reconstruction of the
  primordial curvature power spectrum from PTA data}},
  \href{https://doi.org/10.1088/1475-7516/2024/01/066}{\emph{JCAP} {\bfseries
  01} (2024) 066} [\href{https://arxiv.org/abs/2308.05632}{{\ttfamily
  2308.05632}}].

\bibitem{Ellis:2023oxs}
J.~Ellis, M.~Fairbairn, G.~Franciolini, G.~H\"utsi, A.~Iovino, M.~Lewicki
  et~al., \emph{{What is the source of the PTA GW signal?}},
  \href{https://doi.org/10.1103/PhysRevD.109.023522}{\emph{Phys. Rev. D}
  {\bfseries 109} (2024) 023522}
  [\href{https://arxiv.org/abs/2308.08546}{{\ttfamily 2308.08546}}].

\bibitem{He:2023ado}
S.~He, L.~Li, S.~Wang and S.-J.~Wang, \emph{{Constraints on holographic QCD
  phase transitions from PTA observations}},
  \href{https://arxiv.org/abs/2308.07257}{{\ttfamily 2308.07257}}.

\bibitem{Balaji:2023ehk}
S.~Balaji, G.~Dom\`enech and G.~Franciolini, \emph{{Scalar-induced
  gravitational wave interpretation of PTA data: the role of scalar fluctuation
  propagation speed}},
  \href{https://doi.org/10.1088/1475-7516/2023/10/041}{\emph{JCAP} {\bfseries
  10} (2023) 041} [\href{https://arxiv.org/abs/2307.08552}{{\ttfamily
  2307.08552}}].

\bibitem{Kawasaki:2023rfx}
M.~Kawasaki and K.~Murai, \emph{{Enhancement of gravitational waves at Q-ball
  decay including non-linear density perturbations}},
  \href{https://doi.org/10.1088/1475-7516/2024/01/050}{\emph{JCAP} {\bfseries
  01} (2024) 050} [\href{https://arxiv.org/abs/2308.13134}{{\ttfamily
  2308.13134}}].

\bibitem{Cannizzaro:2023mgc}
E.~Cannizzaro, G.~Franciolini and P.~Pani, \emph{{Novel tests of gravity using
  nano-Hertz stochastic gravitational-wave background signals}},
  \href{https://doi.org/10.1088/1475-7516/2024/04/056}{\emph{JCAP} {\bfseries
  04} (2024) 056} [\href{https://arxiv.org/abs/2307.11665}{{\ttfamily
  2307.11665}}].

\bibitem{King:2023ayw}
S.F.~King, R.~Roshan, X.~Wang, G.~White and M.~Yamazaki, \emph{{Quantum gravity
  effects on dark matter and gravitational waves}},
  \href{https://doi.org/10.1103/PhysRevD.109.024057}{\emph{Phys. Rev. D}
  {\bfseries 109} (2024) 024057}
  [\href{https://arxiv.org/abs/2308.03724}{{\ttfamily 2308.03724}}].

\bibitem{Maji:2023fhv}
R.~Maji and W.-I.~Park, \emph{{Supersymmetric U(1)B-L flat direction and
  NANOGrav 15 year data}},
  \href{https://doi.org/10.1088/1475-7516/2024/01/015}{\emph{JCAP} {\bfseries
  01} (2024) 015} [\href{https://arxiv.org/abs/2308.11439}{{\ttfamily
  2308.11439}}].

\bibitem{Bhaumik:2023wmw}
N.~Bhaumik, R.K.~Jain and M.~Lewicki, \emph{{Ultralow mass primordial black
  holes in the early Universe can explain the pulsar timing array signal}},
  \href{https://doi.org/10.1103/PhysRevD.108.123532}{\emph{Phys. Rev. D}
  {\bfseries 108} (2023) 123532}
  [\href{https://arxiv.org/abs/2308.07912}{{\ttfamily 2308.07912}}].

\bibitem{Zhu:2023lbf}
M.~Zhu, G.~Ye and Y.~Cai, \emph{{Pulsar timing array observations as possible
  hints for nonsingular cosmology}},
  \href{https://doi.org/10.1140/epjc/s10052-023-11963-4}{\emph{Eur. Phys. J. C}
  {\bfseries 83} (2023) 816}
  [\href{https://arxiv.org/abs/2307.16211}{{\ttfamily 2307.16211}}].

\bibitem{Basilakos:2023xof}
S.~Basilakos, D.V.~Nanopoulos, T.~Papanikolaou, E.N.~Saridakis and C.~Tzerefos,
  \emph{{Gravitational wave signatures of no-scale supergravity in NANOGrav and
  beyond}}, \href{https://doi.org/10.1016/j.physletb.2024.138507}{\emph{Phys.
  Lett. B} {\bfseries 850} (2024) 138507}
  [\href{https://arxiv.org/abs/2307.08601}{{\ttfamily 2307.08601}}].

\bibitem{Huang:2023chx}
H.-L.~Huang, Y.~Cai, J.-Q.~Jiang, J.~Zhang and Y.-S.~Piao, \emph{{Supermassive
  primordial black holes in multiverse: for nano-Hertz gravitational wave and
  high-redshift JWST galaxies}},
  \href{https://arxiv.org/abs/2306.17577}{{\ttfamily 2306.17577}}.

\bibitem{Jiang:2023gfe}
J.-Q.~Jiang, Y.~Cai, G.~Ye and Y.-S.~Piao, \emph{{Broken blue-tilted
  inflationary gravitational waves: a joint analysis of NANOGrav 15-year and
  BICEP/Keck 2018 data}},  \href{https://arxiv.org/abs/2307.15547}{{\ttfamily
  2307.15547}}.

\bibitem{DiBari:2023upq}
P.~Di~Bari and M.H.~Rahat, \emph{{The split majoron model confronts the
  NANOGrav signal}},  \href{https://arxiv.org/abs/2307.03184}{{\ttfamily
  2307.03184}}.

\bibitem{Aghaie:2023lan}
M.~Aghaie, G.~Armando, A.~Dondarini and P.~Panci, \emph{{Bounds on Ultralight
  Dark Matter from NANOGrav}},
  \href{https://arxiv.org/abs/2308.04590}{{\ttfamily 2308.04590}}.

\bibitem{Garcia-Saenz:2023zue}
S.~Garcia-Saenz, Y.~Lu and Z.~Shuai, \emph{{Scalar-induced gravitational waves
  from ghost inflation and parity violation}},
  \href{https://doi.org/10.1103/PhysRevD.108.123507}{\emph{Phys. Rev. D}
  {\bfseries 108} (2023) 123507}
  [\href{https://arxiv.org/abs/2306.09052}{{\ttfamily 2306.09052}}].

\bibitem{InternationalPulsarTimingArray:2023mzf}
{\scshape International Pulsar Timing Array} collaboration, \emph{{Comparing
  recent PTA results on the nanohertz stochastic gravitational wave
  background}},  \href{https://arxiv.org/abs/2309.00693}{{\ttfamily
  2309.00693}}.

\bibitem{Ananda:2006af}
K.N.~Ananda, C.~Clarkson and D.~Wands, \emph{{The Cosmological gravitational
  wave background from primordial density perturbations}},
  \href{https://doi.org/10.1103/PhysRevD.75.123518}{\emph{Phys. Rev. D}
  {\bfseries 75} (2007) 123518}
  [\href{https://arxiv.org/abs/gr-qc/0612013}{{\ttfamily gr-qc/0612013}}].

\bibitem{Baumann:2007zm}
D.~Baumann, P.J.~Steinhardt, K.~Takahashi and K.~Ichiki, \emph{{Gravitational
  Wave Spectrum Induced by Primordial Scalar Perturbations}},
  \href{https://doi.org/10.1103/PhysRevD.76.084019}{\emph{Phys. Rev. D}
  {\bfseries 76} (2007) 084019}
  [\href{https://arxiv.org/abs/hep-th/0703290}{{\ttfamily hep-th/0703290}}].

\bibitem{Garcia-Bellido:2016dkw}
J.~Garcia-Bellido, M.~Peloso and C.~Unal, \emph{{Gravitational waves at
  interferometer scales and primordial black holes in axion inflation}},
  \href{https://doi.org/10.1088/1475-7516/2016/12/031}{\emph{JCAP} {\bfseries
  12} (2016) 031} [\href{https://arxiv.org/abs/1610.03763}{{\ttfamily
  1610.03763}}].

\bibitem{Inomata:2016rbd}
K.~Inomata, M.~Kawasaki, K.~Mukaida, Y.~Tada and T.T.~Yanagida,
  \emph{{Inflationary primordial black holes for the LIGO gravitational wave
  events and pulsar timing array experiments}},
  \href{https://doi.org/10.1103/PhysRevD.95.123510}{\emph{Phys. Rev. D}
  {\bfseries 95} (2017) 123510}
  [\href{https://arxiv.org/abs/1611.06130}{{\ttfamily 1611.06130}}].

\bibitem{Garcia-Bellido:2017aan}
J.~Garcia-Bellido, M.~Peloso and C.~Unal, \emph{{Gravitational Wave signatures
  of inflationary models from Primordial Black Hole Dark Matter}},
  \href{https://doi.org/10.1088/1475-7516/2017/09/013}{\emph{JCAP} {\bfseries
  09} (2017) 013} [\href{https://arxiv.org/abs/1707.02441}{{\ttfamily
  1707.02441}}].

\bibitem{Kohri:2018awv}
K.~Kohri and T.~Terada, \emph{{Semianalytic calculation of gravitational wave
  spectrum nonlinearly induced from primordial curvature perturbations}},
  \href{https://doi.org/10.1103/PhysRevD.97.123532}{\emph{Phys. Rev. D}
  {\bfseries 97} (2018) 123532}
  [\href{https://arxiv.org/abs/1804.08577}{{\ttfamily 1804.08577}}].

\bibitem{Cai:2018dig}
R.-g.~Cai, S.~Pi and M.~Sasaki, \emph{{Gravitational Waves Induced by
  non-Gaussian Scalar Perturbations}},
  \href{https://doi.org/10.1103/PhysRevLett.122.201101}{\emph{Phys. Rev. Lett.}
  {\bfseries 122} (2019) 201101}
  [\href{https://arxiv.org/abs/1810.11000}{{\ttfamily 1810.11000}}].

\bibitem{Lu:2019sti}
Y.~Lu, Y.~Gong, Z.~Yi and F.~Zhang, \emph{{Constraints on primordial curvature
  perturbations from primordial black hole dark matter and secondary
  gravitational waves}},
  \href{https://doi.org/10.1088/1475-7516/2019/12/031}{\emph{JCAP} {\bfseries
  12} (2019) 031} [\href{https://arxiv.org/abs/1907.11896}{{\ttfamily
  1907.11896}}].

\bibitem{Yuan:2019wwo}
C.~Yuan, Z.-C.~Chen and Q.-G.~Huang, \emph{{Log-dependent slope of scalar
  induced gravitational waves in the infrared regions}},
  \href{https://doi.org/10.1103/PhysRevD.101.043019}{\emph{Phys. Rev. D}
  {\bfseries 101} (2020) 043019}
  [\href{https://arxiv.org/abs/1910.09099}{{\ttfamily 1910.09099}}].

\bibitem{Chen:2019xse}
Z.-C.~Chen, C.~Yuan and Q.-G.~Huang, \emph{{Pulsar Timing Array Constraints on
  Primordial Black Holes with NANOGrav 11-Year Dataset}},
  \href{https://doi.org/10.1103/PhysRevLett.124.251101}{\emph{Phys. Rev. Lett.}
  {\bfseries 124} (2020) 251101}
  [\href{https://arxiv.org/abs/1910.12239}{{\ttfamily 1910.12239}}].

\bibitem{Xu:2019bdp}
W.-T.~Xu, J.~Liu, T.-J.~Gao and Z.-K.~Guo, \emph{{Gravitational waves from
  double-inflection-point inflation}},
  \href{https://doi.org/10.1103/PhysRevD.101.023505}{\emph{Phys. Rev. D}
  {\bfseries 101} (2020) 023505}
  [\href{https://arxiv.org/abs/1907.05213}{{\ttfamily 1907.05213}}].

\bibitem{Yuan:2019udt}
C.~Yuan, Z.-C.~Chen and Q.-G.~Huang, \emph{{Probing
  primordial\textendash{}black-hole dark matter with scalar induced
  gravitational waves}},
  \href{https://doi.org/10.1103/PhysRevD.100.081301}{\emph{Phys. Rev. D}
  {\bfseries 100} (2019) 081301}
  [\href{https://arxiv.org/abs/1906.11549}{{\ttfamily 1906.11549}}].

\bibitem{Cai:2019cdl}
R.-G.~Cai, S.~Pi and M.~Sasaki, \emph{{Universal infrared scaling of
  gravitational wave background spectra}},
  \href{https://doi.org/10.1103/PhysRevD.102.083528}{\emph{Phys. Rev. D}
  {\bfseries 102} (2020) 083528}
  [\href{https://arxiv.org/abs/1909.13728}{{\ttfamily 1909.13728}}].

\bibitem{Yuan:2019fwv}
C.~Yuan, Z.-C.~Chen and Q.-G.~Huang, \emph{{Scalar induced gravitational waves
  in different gauges}},
  \href{https://doi.org/10.1103/PhysRevD.101.063018}{\emph{Phys. Rev. D}
  {\bfseries 101} (2020) 063018}
  [\href{https://arxiv.org/abs/1912.00885}{{\ttfamily 1912.00885}}].

\bibitem{Yi:2020kmq}
Z.~Yi, Y.~Gong, B.~Wang and Z.-h.~Zhu, \emph{{Primordial black holes and
  secondary gravitational waves from the Higgs field}},
  \href{https://doi.org/10.1103/PhysRevD.103.063535}{\emph{Phys. Rev. D}
  {\bfseries 103} (2021) 063535}
  [\href{https://arxiv.org/abs/2007.09957}{{\ttfamily 2007.09957}}].

\bibitem{Yi:2020cut}
Z.~Yi, Q.~Gao, Y.~Gong and Z.-h.~Zhu, \emph{{Primordial black holes and
  scalar-induced secondary gravitational waves from inflationary models with a
  noncanonical kinetic term}},
  \href{https://doi.org/10.1103/PhysRevD.103.063534}{\emph{Phys. Rev. D}
  {\bfseries 103} (2021) 063534}
  [\href{https://arxiv.org/abs/2011.10606}{{\ttfamily 2011.10606}}].

\bibitem{Liu:2020oqe}
J.~Liu, Z.-K.~Guo and R.-G.~Cai, \emph{{Analytical approximation of the scalar
  spectrum in the ultraslow-roll inflationary models}},
  \href{https://doi.org/10.1103/PhysRevD.101.083535}{\emph{Phys. Rev. D}
  {\bfseries 101} (2020) 083535}
  [\href{https://arxiv.org/abs/2003.02075}{{\ttfamily 2003.02075}}].

\bibitem{Gao:2020tsa}
Q.~Gao, Y.~Gong and Z.~Yi, \emph{{Primordial black holes and secondary
  gravitational waves from natural inflation}},
  \href{https://doi.org/10.1016/j.nuclphysb.2021.115480}{\emph{Nucl. Phys. B}
  {\bfseries 969} (2021) 115480}
  [\href{https://arxiv.org/abs/2012.03856}{{\ttfamily 2012.03856}}].

\bibitem{Yuan:2020iwf}
C.~Yuan and Q.-G.~Huang, \emph{{Gravitational waves induced by the local-type
  non-Gaussian curvature perturbations}},
  \href{https://doi.org/10.1016/j.physletb.2021.136606}{\emph{Phys. Lett. B}
  {\bfseries 821} (2021) 136606}
  [\href{https://arxiv.org/abs/2007.10686}{{\ttfamily 2007.10686}}].

\bibitem{Yuan:2021qgz}
C.~Yuan and Q.-G.~Huang, \emph{{A topic review on probing primordial black hole
  dark matter with scalar induced gravitational waves}},
  \href{https://doi.org/10.1016/j.isci.2021.102860}{\emph{iScience} {\bfseries
  24} (2021) 102860} [\href{https://arxiv.org/abs/2103.04739}{{\ttfamily
  2103.04739}}].

\bibitem{Yi:2021lxc}
Z.~Yi and Z.-H.~Zhu, \emph{{NANOGrav signal and LIGO-Virgo primordial black
  holes from the Higgs field}},
  \href{https://doi.org/10.1088/1475-7516/2022/05/046}{\emph{JCAP} {\bfseries
  05} (2022) 046} [\href{https://arxiv.org/abs/2105.01943}{{\ttfamily
  2105.01943}}].

\bibitem{Yi:2022anu}
Z.~Yi, \emph{{Primordial black holes and scalar-induced gravitational waves
  from the generalized Brans-Dicke theory}},
  \href{https://doi.org/10.1088/1475-7516/2023/03/048}{\emph{JCAP} {\bfseries
  03} (2023) 048} [\href{https://arxiv.org/abs/2206.01039}{{\ttfamily
  2206.01039}}].

\bibitem{Yi:2022ymw}
Z.~Yi and Q.~Fei, \emph{{Constraints on primordial curvature spectrum from
  primordial black holes and scalar-induced gravitational waves}},
  \href{https://doi.org/10.1140/epjc/s10052-023-11233-3}{\emph{Eur. Phys. J. C}
  {\bfseries 83} (2023) 82} [\href{https://arxiv.org/abs/2210.03641}{{\ttfamily
  2210.03641}}].

\bibitem{Yuan:2023ofl}
C.~Yuan, D.-S.~Meng and Q.-G.~Huang, \emph{{Full analysis of the scalar-induced
  gravitational waves for the curvature perturbation with local-type
  non-Gaussianities}},
  \href{https://doi.org/10.1088/1475-7516/2023/12/036}{\emph{JCAP} {\bfseries
  12} (2023) 036} [\href{https://arxiv.org/abs/2308.07155}{{\ttfamily
  2308.07155}}].

\bibitem{Meng:2022ixx}
D.-S.~Meng, C.~Yuan and Q.-g.~Huang, \emph{{One-loop correction to the enhanced
  curvature perturbation with local-type non-Gaussianity for the formation of
  primordial black holes}},
  \href{https://doi.org/10.1103/PhysRevD.106.063508}{\emph{Phys. Rev. D}
  {\bfseries 106} (2022) 063508}
  [\href{https://arxiv.org/abs/2207.07668}{{\ttfamily 2207.07668}}].

\bibitem{Zeldovich:1967lct}
Y.B.~Zel'dovich and I.D.~Novikov, \emph{{The Hypothesis of Cores Retarded
  during Expansion and the Hot Cosmological Model}}, {\emph{Sov. Astron.}
  {\bfseries 10} (1967) 602}.

\bibitem{Hawking:1971ei}
S.~Hawking, \emph{{Gravitationally collapsed objects of very low mass}},
  \href{https://doi.org/10.1093/mnras/152.1.75}{\emph{Mon. Not. Roy. Astron.
  Soc.} {\bfseries 152} (1971) 75}.

\bibitem{Carr:1974nx}
B.J.~Carr and S.W.~Hawking, \emph{{Black holes in the early Universe}},
  \href{https://doi.org/10.1093/mnras/168.2.399}{\emph{Mon. Not. Roy. Astron.
  Soc.} {\bfseries 168} (1974) 399}.

\bibitem{Belotsky:2014kca}
K.M.~Belotsky, A.D.~Dmitriev, E.A.~Esipova, V.A.~Gani, A.V.~Grobov,
  M.Y.~Khlopov et~al., \emph{{Signatures of primordial black hole dark
  matter}}, \href{https://doi.org/10.1142/S0217732314400057}{\emph{Mod. Phys.
  Lett. A} {\bfseries 29} (2014) 1440005}
  [\href{https://arxiv.org/abs/1410.0203}{{\ttfamily 1410.0203}}].

\bibitem{Carr:2016drx}
B.~Carr, F.~Kuhnel and M.~Sandstad, \emph{{Primordial Black Holes as Dark
  Matter}}, \href{https://doi.org/10.1103/PhysRevD.94.083504}{\emph{Phys. Rev.
  D} {\bfseries 94} (2016) 083504}
  [\href{https://arxiv.org/abs/1607.06077}{{\ttfamily 1607.06077}}].

\bibitem{Garcia-Bellido:2017mdw}
J.~Garcia-Bellido and E.~Ruiz~Morales, \emph{{Primordial black holes from
  single field models of inflation}},
  \href{https://doi.org/10.1016/j.dark.2017.09.007}{\emph{Phys. Dark Univ.}
  {\bfseries 18} (2017) 47} [\href{https://arxiv.org/abs/1702.03901}{{\ttfamily
  1702.03901}}].

\bibitem{Carr:2017jsz}
B.~Carr, M.~Raidal, T.~Tenkanen, V.~Vaskonen and H.~Veerm\"ae,
  \emph{{Primordial black hole constraints for extended mass functions}},
  \href{https://doi.org/10.1103/PhysRevD.96.023514}{\emph{Phys. Rev. D}
  {\bfseries 96} (2017) 023514}
  [\href{https://arxiv.org/abs/1705.05567}{{\ttfamily 1705.05567}}].

\bibitem{Germani:2017bcs}
C.~Germani and T.~Prokopec, \emph{{On primordial black holes from an inflection
  point}}, \href{https://doi.org/10.1016/j.dark.2017.09.001}{\emph{Phys. Dark
  Univ.} {\bfseries 18} (2017) 6}
  [\href{https://arxiv.org/abs/1706.04226}{{\ttfamily 1706.04226}}].

\bibitem{Chen:2018rzo}
Z.-C.~Chen, F.~Huang and Q.-G.~Huang, \emph{{Stochastic Gravitational-wave
  Background from Binary Black Holes and Binary Neutron Stars and Implications
  for LISA}}, \href{https://doi.org/10.3847/1538-4357/aaf581}{\emph{Astrophys.
  J.} {\bfseries 871} (2019) 97}
  [\href{https://arxiv.org/abs/1809.10360}{{\ttfamily 1809.10360}}].

\bibitem{Liu:2018ess}
L.~Liu, Z.-K.~Guo and R.-G.~Cai, \emph{{Effects of the surrounding primordial
  black holes on the merger rate of primordial black hole binaries}},
  \href{https://doi.org/10.1103/PhysRevD.99.063523}{\emph{Phys. Rev. D}
  {\bfseries 99} (2019) 063523}
  [\href{https://arxiv.org/abs/1812.05376}{{\ttfamily 1812.05376}}].

\bibitem{Chen:2018czv}
Z.-C.~Chen and Q.-G.~Huang, \emph{{Merger Rate Distribution of
  Primordial-Black-Hole Binaries}},
  \href{https://doi.org/10.3847/1538-4357/aad6e2}{\emph{Astrophys. J.}
  {\bfseries 864} (2018) 61}
  [\href{https://arxiv.org/abs/1801.10327}{{\ttfamily 1801.10327}}].

\bibitem{Liu:2019rnx}
L.~Liu, Z.-K.~Guo and R.-G.~Cai, \emph{{Effects of the merger history on the
  merger rate density of primordial black hole binaries}},
  \href{https://doi.org/10.1140/epjc/s10052-019-7227-0}{\emph{Eur. Phys. J. C}
  {\bfseries 79} (2019) 717}
  [\href{https://arxiv.org/abs/1901.07672}{{\ttfamily 1901.07672}}].

\bibitem{Fu:2019ttf}
C.~Fu, P.~Wu and H.~Yu, \emph{{Primordial Black Holes from Inflation with
  Nonminimal Derivative Coupling}},
  \href{https://doi.org/10.1103/PhysRevD.100.063532}{\emph{Phys. Rev. D}
  {\bfseries 100} (2019) 063532}
  [\href{https://arxiv.org/abs/1907.05042}{{\ttfamily 1907.05042}}].

\bibitem{Liu:2019lul}
J.~Liu, Z.-K.~Guo and R.-G.~Cai, \emph{{Primordial Black Holes from Cosmic
  Domain Walls}},
  \href{https://doi.org/10.1103/PhysRevD.101.023513}{\emph{Phys. Rev. D}
  {\bfseries 101} (2020) 023513}
  [\href{https://arxiv.org/abs/1908.02662}{{\ttfamily 1908.02662}}].

\bibitem{Cai:2019bmk}
R.-G.~Cai, Z.-K.~Guo, J.~Liu, L.~Liu and X.-Y.~Yang, \emph{{Primordial black
  holes and gravitational waves from parametric amplification of curvature
  perturbations}},
  \href{https://doi.org/10.1088/1475-7516/2020/06/013}{\emph{JCAP} {\bfseries
  06} (2020) 013} [\href{https://arxiv.org/abs/1912.10437}{{\ttfamily
  1912.10437}}].

\bibitem{Chen:2019irf}
Z.-C.~Chen and Q.-G.~Huang, \emph{{Distinguishing Primordial Black Holes from
  Astrophysical Black Holes by Einstein Telescope and Cosmic Explorer}},
  \href{https://doi.org/10.1088/1475-7516/2020/08/039}{\emph{JCAP} {\bfseries
  08} (2020) 039} [\href{https://arxiv.org/abs/1904.02396}{{\ttfamily
  1904.02396}}].

\bibitem{Liu:2020cds}
L.~Liu, Z.-K.~Guo, R.-G.~Cai and S.P.~Kim, \emph{{Merger rate distribution of
  primordial black hole binaries with electric charges}},
  \href{https://doi.org/10.1103/PhysRevD.102.043508}{\emph{Phys. Rev. D}
  {\bfseries 102} (2020) 043508}
  [\href{https://arxiv.org/abs/2001.02984}{{\ttfamily 2001.02984}}].

\bibitem{Fu:2020lob}
C.~Fu, P.~Wu and H.~Yu, \emph{{Primordial black holes and oscillating
  gravitational waves in slow-roll and slow-climb inflation with an
  intermediate noninflationary phase}},
  \href{https://doi.org/10.1103/PhysRevD.102.043527}{\emph{Phys. Rev. D}
  {\bfseries 102} (2020) 043527}
  [\href{https://arxiv.org/abs/2006.03768}{{\ttfamily 2006.03768}}].

\bibitem{Liu:2020vsy}
L.~Liu, O.~Christiansen, Z.-K.~Guo, R.-G.~Cai and S.P.~Kim,
  \emph{{Gravitational and electromagnetic radiation from binary black holes
  with electric and magnetic charges: Circular orbits on a cone}},
  \href{https://doi.org/10.1103/PhysRevD.102.103520}{\emph{Phys. Rev. D}
  {\bfseries 102} (2020) 103520}
  [\href{https://arxiv.org/abs/2008.02326}{{\ttfamily 2008.02326}}].

\bibitem{Liu:2020bag}
L.~Liu, O.~Christiansen, W.-H.~Ruan, Z.-K.~Guo, R.-G.~Cai and S.P.~Kim,
  \emph{{Gravitational and electromagnetic radiation from binary black holes
  with electric and magnetic charges: elliptical orbits on a cone}},
  \href{https://doi.org/10.1140/epjc/s10052-021-09849-4}{\emph{Eur. Phys. J. C}
  {\bfseries 81} (2021) 1048}
  [\href{https://arxiv.org/abs/2011.13586}{{\ttfamily 2011.13586}}].

\bibitem{Wu:2020drm}
Y.~Wu, \emph{{Merger history of primordial black-hole binaries}},
  \href{https://doi.org/10.1103/PhysRevD.101.083008}{\emph{Phys. Rev. D}
  {\bfseries 101} (2020) 083008}
  [\href{https://arxiv.org/abs/2001.03833}{{\ttfamily 2001.03833}}].

\bibitem{DeLuca:2020sae}
V.~De~Luca, V.~Desjacques, G.~Franciolini, P.~Pani and A.~Riotto,
  \emph{{GW190521 Mass Gap Event and the Primordial Black Hole Scenario}},
  \href{https://doi.org/10.1103/PhysRevLett.126.051101}{\emph{Phys. Rev. Lett.}
  {\bfseries 126} (2021) 051101}
  [\href{https://arxiv.org/abs/2009.01728}{{\ttfamily 2009.01728}}].

\bibitem{Vaskonen:2020lbd}
V.~Vaskonen and H.~Veerm\"ae, \emph{{Did NANOGrav see a signal from primordial
  black hole formation?}},
  \href{https://doi.org/10.1103/PhysRevLett.126.051303}{\emph{Phys. Rev. Lett.}
  {\bfseries 126} (2021) 051303}
  [\href{https://arxiv.org/abs/2009.07832}{{\ttfamily 2009.07832}}].

\bibitem{DeLuca:2020agl}
V.~De~Luca, G.~Franciolini and A.~Riotto, \emph{{NANOGrav Data Hints at
  Primordial Black Holes as Dark Matter}},
  \href{https://doi.org/10.1103/PhysRevLett.126.041303}{\emph{Phys. Rev. Lett.}
  {\bfseries 126} (2021) 041303}
  [\href{https://arxiv.org/abs/2009.08268}{{\ttfamily 2009.08268}}].

\bibitem{Domenech:2020ers}
G.~Dom\`enech and S.~Pi, \emph{{NANOGrav hints on planet-mass primordial black
  holes}}, \href{https://doi.org/10.1007/s11433-021-1839-6}{\emph{Sci. China
  Phys. Mech. Astron.} {\bfseries 65} (2022) 230411}
  [\href{https://arxiv.org/abs/2010.03976}{{\ttfamily 2010.03976}}].

\bibitem{Hutsi:2020sol}
G.~H\"utsi, M.~Raidal, V.~Vaskonen and H.~Veerm\"ae, \emph{{Two populations of
  LIGO-Virgo black holes}},
  \href{https://doi.org/10.1088/1475-7516/2021/03/068}{\emph{JCAP} {\bfseries
  03} (2021) 068} [\href{https://arxiv.org/abs/2012.02786}{{\ttfamily
  2012.02786}}].

\bibitem{Chen:2021nxo}
Z.-C.~Chen, C.~Yuan and Q.-G.~Huang, \emph{{Confronting the primordial black
  hole scenario with the gravitational-wave events detected by LIGO-Virgo}},
  \href{https://doi.org/10.1016/j.physletb.2022.137040}{\emph{Phys. Lett. B}
  {\bfseries 829} (2022) 137040}
  [\href{https://arxiv.org/abs/2108.11740}{{\ttfamily 2108.11740}}].

\bibitem{Kawai:2021edk}
S.~Kawai and J.~Kim, \emph{{Primordial black holes from Gauss-Bonnet-corrected
  single field inflation}},
  \href{https://doi.org/10.1103/PhysRevD.104.083545}{\emph{Phys. Rev. D}
  {\bfseries 104} (2021) 083545}
  [\href{https://arxiv.org/abs/2108.01340}{{\ttfamily 2108.01340}}].

\bibitem{Braglia:2021wwa}
M.~Braglia, J.~Garcia-Bellido and S.~Kuroyanagi, \emph{{Testing Primordial
  Black Holes with multi-band observations of the stochastic gravitational wave
  background}},
  \href{https://doi.org/10.1088/1475-7516/2021/12/012}{\emph{JCAP} {\bfseries
  12} (2021) 012} [\href{https://arxiv.org/abs/2110.07488}{{\ttfamily
  2110.07488}}].

\bibitem{Cai:2021wzd}
R.-G.~Cai, C.~Chen and C.~Fu, \emph{{Primordial black holes and stochastic
  gravitational wave background from inflation with a noncanonical spectator
  field}}, \href{https://doi.org/10.1103/PhysRevD.104.083537}{\emph{Phys. Rev.
  D} {\bfseries 104} (2021) 083537}
  [\href{https://arxiv.org/abs/2108.03422}{{\ttfamily 2108.03422}}].

\bibitem{Liu:2021jnw}
L.~Liu, X.-Y.~Yang, Z.-K.~Guo and R.-G.~Cai, \emph{{Testing primordial black
  hole and measuring the Hubble constant with multiband gravitational-wave
  observations}},
  \href{https://doi.org/10.1088/1475-7516/2023/01/006}{\emph{JCAP} {\bfseries
  01} (2023) 006} [\href{https://arxiv.org/abs/2112.05473}{{\ttfamily
  2112.05473}}].

\bibitem{Braglia:2022icu}
M.~Braglia, J.~Garcia-Bellido and S.~Kuroyanagi, \emph{{Tracking the origin of
  black holes with the stochastic gravitational wave background popcorn
  signal}}, \href{https://doi.org/10.1093/mnras/stad082}{\emph{Mon. Not. Roy.
  Astron. Soc.} {\bfseries 519} (2023) 6008}
  [\href{https://arxiv.org/abs/2201.13414}{{\ttfamily 2201.13414}}].

\bibitem{Liu:2022wtq}
L.~Liu and S.P.~Kim, \emph{{Merger rate of charged black holes from the
  two-body dynamical capture}},
  \href{https://doi.org/10.1088/1475-7516/2022/03/059}{\emph{JCAP} {\bfseries
  03} (2022) 059} [\href{https://arxiv.org/abs/2201.02581}{{\ttfamily
  2201.02581}}].

\bibitem{Zheng:2022wqo}
L.-M.~Zheng, Z.~Li, Z.-C.~Chen, H.~Zhou and Z.-H.~Zhu, \emph{{Towards a
  reliable reconstruction of the power spectrum of primordial curvature
  perturbation on small scales from GWTC-3}},
  \href{https://doi.org/10.1016/j.physletb.2023.137720}{\emph{Phys. Lett. B}
  {\bfseries 838} (2023) 137720}
  [\href{https://arxiv.org/abs/2212.05516}{{\ttfamily 2212.05516}}].

\bibitem{Chen:2022qvg}
Z.-C.~Chen, S.P.~Kim and L.~Liu, \emph{{Gravitational and electromagnetic
  radiation from binary black holes with electric and magnetic charges:
  hyperbolic orbits on a cone}},
  \href{https://doi.org/10.1088/1572-9494/acce98}{\emph{Commun. Theor. Phys.}
  {\bfseries 75} (2023) 065401}
  [\href{https://arxiv.org/abs/2210.15564}{{\ttfamily 2210.15564}}].

\bibitem{Liu:2022iuf}
L.~Liu, Z.-Q.~You, Y.~Wu and Z.-C.~Chen, \emph{{Constraining the merger history
  of primordial-black-hole binaries from GWTC-3}},
  \href{https://doi.org/10.1103/PhysRevD.107.063035}{\emph{Phys. Rev. D}
  {\bfseries 107} (2023) 063035}
  [\href{https://arxiv.org/abs/2210.16094}{{\ttfamily 2210.16094}}].

\bibitem{Chen:2022fda}
Z.-C.~Chen, S.-S.~Du, Q.-G.~Huang and Z.-Q.~You, \emph{{Constraints on
  primordial-black-hole population and cosmic expansion history from GWTC-3}},
  \href{https://doi.org/10.1088/1475-7516/2023/03/024}{\emph{JCAP} {\bfseries
  03} (2023) 024} [\href{https://arxiv.org/abs/2205.11278}{{\ttfamily
  2205.11278}}].

\bibitem{Inomata:2022yte}
K.~Inomata, M.~Braglia, X.~Chen and S.~Renaux-Petel, \emph{{Questions on
  calculation of primordial power spectrum with large spikes: the resonance
  model case}},
  \href{https://doi.org/10.1088/1475-7516/2023/04/011}{\emph{JCAP} {\bfseries
  04} (2023) 011} [\href{https://arxiv.org/abs/2211.02586}{{\ttfamily
  2211.02586}}].

\bibitem{Guo:2023hyp}
S.-Y.~Guo, M.~Khlopov, X.~Liu, L.~Wu, Y.~Wu and B.~Zhu, \emph{{Footprints of
  Axion-Like Particle in Pulsar Timing Array Data and JWST Observations}},
  \href{https://arxiv.org/abs/2306.17022}{{\ttfamily 2306.17022}}.

\bibitem{Cai:2023uhc}
Y.~Cai, M.~Zhu and Y.-S.~Piao, \emph{{Primordial black holes from null energy
  condition violation during inflation}},
  \href{https://arxiv.org/abs/2305.10933}{{\ttfamily 2305.10933}}.

\bibitem{Meng:2022low}
D.-S.~Meng, C.~Yuan and Q.-G.~Huang, \emph{{Primordial black holes generated by
  the non-minimal spectator field}},
  \href{https://doi.org/10.1007/s11433-022-2095-5}{\emph{Sci. China Phys. Mech.
  Astron.} {\bfseries 66} (2023) 280411}
  [\href{https://arxiv.org/abs/2212.03577}{{\ttfamily 2212.03577}}].

\bibitem{Sasaki:2018dmp}
M.~Sasaki, T.~Suyama, T.~Tanaka and S.~Yokoyama, \emph{{Primordial black
  holes\textemdash{}perspectives in gravitational wave astronomy}},
  \href{https://doi.org/10.1088/1361-6382/aaa7b4}{\emph{Class. Quant. Grav.}
  {\bfseries 35} (2018) 063001}
  [\href{https://arxiv.org/abs/1801.05235}{{\ttfamily 1801.05235}}].

\bibitem{Carr:2020gox}
B.~Carr, K.~Kohri, Y.~Sendouda and J.~Yokoyama, \emph{{Constraints on
  primordial black holes}},
  \href{https://doi.org/10.1088/1361-6633/ac1e31}{\emph{Rept. Prog. Phys.}
  {\bfseries 84} (2021) 116902}
  [\href{https://arxiv.org/abs/2002.12778}{{\ttfamily 2002.12778}}].

\bibitem{Carr:2020xqk}
B.~Carr and F.~Kuhnel, \emph{{Primordial Black Holes as Dark Matter: Recent
  Developments}},
  \href{https://doi.org/10.1146/annurev-nucl-050520-125911}{\emph{Ann. Rev.
  Nucl. Part. Sci.} {\bfseries 70} (2020) 355}
  [\href{https://arxiv.org/abs/2006.02838}{{\ttfamily 2006.02838}}].

\bibitem{Bird:2016dcv}
S.~Bird, I.~Cholis, J.B.~Mu\~noz, Y.~Ali-Ha\"\i{}moud, M.~Kamionkowski,
  E.D.~Kovetz et~al., \emph{{Did LIGO detect dark matter?}},
  \href{https://doi.org/10.1103/PhysRevLett.116.201301}{\emph{Phys. Rev. Lett.}
  {\bfseries 116} (2016) 201301}
  [\href{https://arxiv.org/abs/1603.00464}{{\ttfamily 1603.00464}}].

\bibitem{Sasaki:2016jop}
M.~Sasaki, T.~Suyama, T.~Tanaka and S.~Yokoyama, \emph{{Primordial Black Hole
  Scenario for the Gravitational-Wave Event GW150914}},
  \href{https://doi.org/10.1103/PhysRevLett.117.061101}{\emph{Phys. Rev. Lett.}
  {\bfseries 117} (2016) 061101}
  [\href{https://arxiv.org/abs/1603.08338}{{\ttfamily 1603.08338}}].

\bibitem{Scholtz:2019csj}
J.~Scholtz and J.~Unwin, \emph{{What if Planet 9 is a Primordial Black Hole?}},
  \href{https://doi.org/10.1103/PhysRevLett.125.051103}{\emph{Phys. Rev. Lett.}
  {\bfseries 125} (2020) 051103}
  [\href{https://arxiv.org/abs/1909.11090}{{\ttfamily 1909.11090}}].

\bibitem{Mroz:2017mvf}
P.~Mroz et~al., \emph{{No large population of unbound or wide-orbit
  Jupiter-mass planets}},
  \href{https://doi.org/10.1038/nature23276}{\emph{Nature} {\bfseries 548}
  (2017) 183} [\href{https://arxiv.org/abs/1707.07634}{{\ttfamily
  1707.07634}}].

\bibitem{Niikura:2019kqi}
H.~Niikura, M.~Takada, S.~Yokoyama, T.~Sumi and S.~Masaki, \emph{{Constraints
  on Earth-mass primordial black holes from OGLE 5-year microlensing events}},
  \href{https://doi.org/10.1103/PhysRevD.99.083503}{\emph{Phys. Rev. D}
  {\bfseries 99} (2019) 083503}
  [\href{https://arxiv.org/abs/1901.07120}{{\ttfamily 1901.07120}}].

\bibitem{Kodama:1984ziu}
H.~Kodama and M.~Sasaki, \emph{{Cosmological Perturbation Theory}},
  \href{https://doi.org/10.1143/PTPS.78.1}{\emph{Prog. Theor. Phys. Suppl.}
  {\bfseries 78} (1984) 1}.

\bibitem{Mukhanov:1990me}
V.F.~Mukhanov, H.A.~Feldman and R.H.~Brandenberger, \emph{{Theory of
  cosmological perturbations. Part 1. Classical perturbations. Part 2. Quantum
  theory of perturbations. Part 3. Extensions}},
  \href{https://doi.org/10.1016/0370-1573(92)90044-Z}{\emph{Phys. Rept.}
  {\bfseries 215} (1992) 203}.

\bibitem{Bezrukov:2007ep}
F.L.~Bezrukov and M.~Shaposhnikov, \emph{{The Standard Model Higgs boson as the
  inflaton}}, \href{https://doi.org/10.1016/j.physletb.2007.11.072}{\emph{Phys.
  Lett. B} {\bfseries 659} (2008) 703}
  [\href{https://arxiv.org/abs/0710.3755}{{\ttfamily 0710.3755}}].

\bibitem{Hwang:1996bc}
J.-c.~Hwang, \emph{{Quantum fluctuations of cosmological perturbations in
  generalized gravity}},
  \href{https://doi.org/10.1088/0264-9381/14/12/016}{\emph{Class. Quant. Grav.}
  {\bfseries 14} (1997) 3327}
  [\href{https://arxiv.org/abs/gr-qc/9607059}{{\ttfamily gr-qc/9607059}}].

\bibitem{Weinberg:2003ur}
S.~Weinberg, \emph{{Damping of tensor modes in cosmology}},
  \href{https://doi.org/10.1103/PhysRevD.69.023503}{\emph{Phys. Rev. D}
  {\bfseries 69} (2004) 023503}
  [\href{https://arxiv.org/abs/astro-ph/0306304}{{\ttfamily
  astro-ph/0306304}}].

\bibitem{Watanabe:2006qe}
Y.~Watanabe and E.~Komatsu, \emph{{Improved Calculation of the Primordial
  Gravitational Wave Spectrum in the Standard Model}},
  \href{https://doi.org/10.1103/PhysRevD.73.123515}{\emph{Phys. Rev. D}
  {\bfseries 73} (2006) 123515}
  [\href{https://arxiv.org/abs/astro-ph/0604176}{{\ttfamily
  astro-ph/0604176}}].

\bibitem{Espinosa:2018eve}
J.R.~Espinosa, D.~Racco and A.~Riotto, \emph{{A Cosmological Signature of the
  SM Higgs Instability: Gravitational Waves}},
  \href{https://doi.org/10.1088/1475-7516/2018/09/012}{\emph{JCAP} {\bfseries
  09} (2018) 012} [\href{https://arxiv.org/abs/1804.07732}{{\ttfamily
  1804.07732}}].

\bibitem{Inomata:2018cht}
K.~Inomata, M.~Kawasaki, K.~Mukaida and T.T.~Yanagida, \emph{{Double inflation
  as a single origin of primordial black holes for all dark matter and LIGO
  observations}}, \href{https://doi.org/10.1103/PhysRevD.97.043514}{\emph{Phys.
  Rev. D} {\bfseries 97} (2018) 043514}
  [\href{https://arxiv.org/abs/1711.06129}{{\ttfamily 1711.06129}}].

\bibitem{Planck:2018jri}
{\scshape Planck} collaboration, \emph{{Planck 2018 results. X. Constraints on
  inflation}}, \href{https://doi.org/10.1051/0004-6361/201833887}{\emph{Astron.
  Astrophys.} {\bfseries 641} (2020) A10}
  [\href{https://arxiv.org/abs/1807.06211}{{\ttfamily 1807.06211}}].

\bibitem{Inomata:2016uip}
K.~Inomata, M.~Kawasaki and Y.~Tada, \emph{{Revisiting constraints on small
  scale perturbations from big-bang nucleosynthesis}},
  \href{https://doi.org/10.1103/PhysRevD.94.043527}{\emph{Phys. Rev. D}
  {\bfseries 94} (2016) 043527}
  [\href{https://arxiv.org/abs/1605.04646}{{\ttfamily 1605.04646}}].

\bibitem{Jeong:2014gna}
D.~Jeong, J.~Pradler, J.~Chluba and M.~Kamionkowski, \emph{{Silk damping at a
  redshift of a billion: a new limit on small-scale adiabatic perturbations}},
  \href{https://doi.org/10.1103/PhysRevLett.113.061301}{\emph{Phys. Rev. Lett.}
  {\bfseries 113} (2014) 061301}
  [\href{https://arxiv.org/abs/1403.3697}{{\ttfamily 1403.3697}}].

\bibitem{Fixsen:1996nj}
D.J.~Fixsen, E.S.~Cheng, J.M.~Gales, J.C.~Mather, R.A.~Shafer and E.L.~Wright,
  \emph{{The Cosmic Microwave Background spectrum from the full COBE FIRAS data
  set}}, \href{https://doi.org/10.1086/178173}{\emph{Astrophys. J.} {\bfseries
  473} (1996) 576} [\href{https://arxiv.org/abs/astro-ph/9605054}{{\ttfamily
  astro-ph/9605054}}].

\bibitem{Chluba:2012we}
J.~Chluba, A.L.~Erickcek and I.~Ben-Dayan, \emph{{Probing the inflaton:
  Small-scale power spectrum constraints from measurements of the CMB energy
  spectrum}},
  \href{https://doi.org/10.1088/0004-637X/758/2/76}{\emph{Astrophys. J.}
  {\bfseries 758} (2012) 76} [\href{https://arxiv.org/abs/1203.2681}{{\ttfamily
  1203.2681}}].

\bibitem{Theia:2017xtk}
{\scshape Theia} collaboration, \emph{{Theia: Faint objects in motion or the
  new astrometry frontier}},
  \href{https://arxiv.org/abs/1707.01348}{{\ttfamily 1707.01348}}.

\bibitem{Sesana:2019vho}
A.~Sesana et~al., \emph{{Unveiling the gravitational universe at $\mu$-Hz
  frequencies}}, \href{https://doi.org/10.1007/s10686-021-09709-9}{\emph{Exper.
  Astron.} {\bfseries 51} (2021) 1333}
  [\href{https://arxiv.org/abs/1908.11391}{{\ttfamily 1908.11391}}].

\bibitem{EROS-2:2006ryy}
{\scshape EROS-2} collaboration, \emph{{Limits on the Macho Content of the
  Galactic Halo from the EROS-2 Survey of the Magellanic Clouds}},
  \href{https://doi.org/10.1051/0004-6361:20066017}{\emph{Astron. Astrophys.}
  {\bfseries 469} (2007) 387}
  [\href{https://arxiv.org/abs/astro-ph/0607207}{{\ttfamily
  astro-ph/0607207}}].

\bibitem{Niikura:2017zjd}
H.~Niikura et~al., \emph{{Microlensing constraints on primordial black holes
  with Subaru/HSC Andromeda observations}},
  \href{https://doi.org/10.1038/s41550-019-0723-1}{\emph{Nature Astron.}
  {\bfseries 3} (2019) 524} [\href{https://arxiv.org/abs/1701.02151}{{\ttfamily
  1701.02151}}].

\bibitem{Griest:2013esa}
K.~Griest, A.M.~Cieplak and M.J.~Lehner, \emph{{New Limits on Primordial Black
  Hole Dark Matter from an Analysis of Kepler Source Microlensing Data}},
  \href{https://doi.org/10.1103/PhysRevLett.111.181302}{\emph{Phys. Rev. Lett.}
  {\bfseries 111} (2013) 181302}.

\bibitem{Moore:2021ibq}
C.J.~Moore and A.~Vecchio, \emph{{Ultra-low-frequency gravitational waves from
  cosmological and astrophysical processes}},
  \href{https://doi.org/10.1038/s41550-021-01489-8}{\emph{Nature Astron.}
  {\bfseries 5} (2021) 1268}
  [\href{https://arxiv.org/abs/2104.15130}{{\ttfamily 2104.15130}}].

\bibitem{Janssen:2014dka}
G.~Janssen et~al., \emph{{Gravitational wave astronomy with the SKA}},
  \href{https://doi.org/10.22323/1.215.0037}{\emph{PoS} {\bfseries AASKA14}
  (2015) 037} [\href{https://arxiv.org/abs/1501.00127}{{\ttfamily
  1501.00127}}].

\bibitem{Kusenko:2020pcg}
A.~Kusenko, M.~Sasaki, S.~Sugiyama, M.~Takada, V.~Takhistov and E.~Vitagliano,
  \emph{{Exploring Primordial Black Holes from the Multiverse with Optical
  Telescopes}},
  \href{https://doi.org/10.1103/PhysRevLett.125.181304}{\emph{Phys. Rev. Lett.}
  {\bfseries 125} (2020) 181304}
  [\href{https://arxiv.org/abs/2001.09160}{{\ttfamily 2001.09160}}].

\bibitem{Siraj:2020upy}
A.~Siraj and A.~Loeb, \emph{{Searching for Black Holes in the Outer Solar
  System with LSST}},
  \href{https://doi.org/10.3847/2041-8213/aba119}{\emph{Astrophys. J. Lett.}
  {\bfseries 898} (2020) L4}
  [\href{https://arxiv.org/abs/2005.12280}{{\ttfamily 2005.12280}}].

\bibitem{Arbey:2020urq}
A.~Arbey and J.~Auffinger, \emph{{Detecting Planet 9 via Hawking radiation}},
  \href{https://arxiv.org/abs/2006.02944}{{\ttfamily 2006.02944}}.

\bibitem{Witten:2020ifl}
E.~Witten, \emph{{Searching for a Black Hole in the Outer Solar System}},
  \href{https://arxiv.org/abs/2004.14192}{{\ttfamily 2004.14192}}.

\end{thebibliography}\endgroup
\end{document}